\documentclass[a4paper]{amsart}

\usepackage{mathpple}
\usepackage[a4paper]{geometry}
\usepackage[latin1]{inputenc}
\usepackage[british]{babel}
\usepackage{graphicx}
\graphicspath{{graphics/}}

\RequirePackage{mathrsfs}
\DeclareSymbolFontAlphabet{\mathrsfs}{rsfs}

\usepackage{tensind}
\tensordelimiter{°}

\RequirePackage[newenum]{paralist}
\numberwithin{equation}{section}

\usepackage{url}

\newcommand{\cC}{{\mathcal C}}         
\newcommand{\cD}{{\mathcal D}}

\newcommand{\cP}{{\mathcal P}}

\newcommand{\cZ}{{\mathcal Z}}         

\newbox\idbox

\def\mathds{\mathbb}

\newcommand{\name}[1]{#1}

\newtheorem{theorem}{Theorem}

\newcommand{\pr}[1]{\ref{#1}}

\newcommand{\eqr}[1]{\eqref{eq:#1}}

\newcommand{\eql}[1]{\label{eq:#1}}

\newcommand{\figr}[1]{fig.~\ref{fig:#1}}

\DeclareMathOperator{\sign}{sign}

\renewcommand{\epsilon}{\varepsilon}

\newcommand{\us}{\underset}

\newcommand{\tf}{\tfrac}

\newcommand{\gdw}{\quad\Leftrightarrow\quad}

\newcommand{\imp}{\Rightarrow}

\newcommand{\falle}{\quad\forall\,}

\newcommand{\ntext}[1]{\quad\text{#1}}

\newcommand{\ztext}[1]{\quad\text{#1}\quad}

\newcommand{\N}{\mathds{N}}

\newcommand{\eps}{\epsilon}
\newcommand{\del}{\partial}

\newcommand{\E}{\mathcal{E}}%
\newcommand{\B}{\mathcal{B}}%
\newcommand{\F}{\mathcal{F}}%
\newcommand{\X}{\mathcal{X}}%

\newcommand{\chiC}{\mathcal{K}}%

\newcommand{\mf}{\mathcal{M}}

\newcommand{\Fh}{\widehat\F}
\renewcommand{\P}{\mathcal{P}}
\newcommand{\M}{\mathcal{M}}
\newcommand{\Mt}{{\widetilde\M}}
\newcommand{\sN}{\mathcal{N}}
\newcommand{\isN}{{\widetilde\sN}}

\newcommand{\pathC}{\mathrsfs{C}}

\newcommand{\hd}[1]{\cD_{#1}}

\newcommand{\trez}[1]{\tfrac{1}{#1}}
\newcommand{\half}{\trez{2}}

\newcommand{\non}{\nonumber}

\newcommand{\ra}{\rightarrow}

\newcommand{\ceq}{\displaybreak[0]\\ \non &}
\newcommand{\beq}{\displaybreak[0]\\}
\newcommand{\nbeq}{\displaybreak[0]\\ \non}
\newcommand{\cbeq}{\,,\beq}
\newcommand{\cnbeq}{\,,\nbeq}
\newcommand{\ct}{\\&\rule{\eqspace}{0pt}\non}

\newcommand{\cp}[1]{\X_#1}

\whenindex{A}{a'}{}
\whenindex{B}{b'}{}
\whenindex{C}{c'}{}
\whenindex{D}{d'}{}
\whenindex{E}{e'}{}

\whenindex{T}{\bullet}{}

\newlength{\amdw}
\newlength{\aml}
\newlength{\amr}
\newlength{\amrgi}
\newlength{\amrgii}

\newcommand{\alignmiddle}[2]{%
  \settowidth{\aml}{$#1$} %
  \hspace{\aml} %
  &
  \setlength{\fboxsep}{0pt} %
  \settowidth{\amdw}{$.$} %
  \settowidth{\amr}{$#2$} %
  \settowidth{\amrgi}{$.#2$} %
  \addtolength{\amrgi}{-\amdw} %
  \addtolength{\amrgi}{-\amr} %
  \settowidth{\amrgii}{$#2.$} %
  \addtolength{\amrgii}{-\amdw} %
  \addtolength{\amrgii}{-\amr} %
  \hspace*{\amrgi}\makebox[\amr][r]{$#1#2$}\hspace*{\amrgii}
  }

\newcommand{\am}{\alignmiddle}

\newlength{\stms}

\newcommand{\settomathskip}[3]{
  \settowidth{#1}{$#2#3$}
  \settowidth{\stms}{$#3$}
  \addtolength{#1}{-\stms}
  }

\newlength{\eqspace}
\settomathskip{\eqspace}{=}{-}
\newcommand{\rh}[1]{\eqr{rh#1}}

\begin{document}

\title{ Algebraic stability analysis of constraint propagation }

\author{J.~Frauendiener}
\address[J.~Frauendiener]{Institut für Astronomie und Astrophysik\\
  Universität Tübingen\\
  Auf der Morgenstelle 10\\
  72076 Tübingen\\
  Germany}
\email{joerg.frauendiener@uni-tuebingen.de}

\author{T.~Vogel}
\address[T.~Vogel]{Institut für Astronomie und Astrophysik\\
  Universität Tübingen\\
  Auf der Morgenstelle 10\\
  72076 Tübingen\\
  Germany;
Max-Planck-Institut für~Gravitationsphysik
  (Albert-Einstein-Institut) \\
  Am~Mühlenberg~1 \\
  14476~Golm \\
  Germany}
\email{tilman.vogel@aei.mpg.de}

\thanks{Publication numbers: AEI-2004-091, ESI-1531\\
  \indent
  Journal reference: Class. Quantum Grav. \textbf{22} (2005) 1769-1793\\
  \indent
  © copyright (2005) IOP Publishing Ltd.,
  \url{http://www.iop.org/}}
\begin{abstract}
  The divergence of the constraint quantities is a major problem in
  computational gravity today. Apparently, there are two sources for
  constraint violations. The use of boundary conditions which are not
  compatible with the constraint equations inadvertently leads to
  `constraint violating modes' propagating into the computational
  domain from the boundary. The other source for constraint violation
  is intrinsic. It is already present in the initial value problem,
  i.e. even when no boundary conditions have to be specified. Its
  origin is due to the instability of the constraint surface in the
  phase space of initial conditions for the time evolution equations.
  In this paper, we present a technique to study in detail how this
  instability depends on gauge parameters. We demonstrate this for the
  influence of the choice of the time foliation in context of the Weyl
  system. This system is the essential hyperbolic part in various
  formulations of the Einstein equations.
\end{abstract}

\maketitle


\begin{lrbox}{\idbox}
\verb$Id: introduction.tex,v 1.19 2005/07/21 12:51:08 vogel Rel $
\end{lrbox}


\def\cP{\mathcal{P}}
\def\cC{\mathcal{C}}
\def\cZ{\mathcal{Z}}

\section{Introduction}
\label{sec:intro}

\noindent
One of the major problems in computational gravity is the fact that
the constraints are not preserved in free evolution codes. Indeed, it
can be observed in many numerical approaches that the constraints are
violated with an exponential (or even worse) rate in time. Thus, the
numerically generated solution of the evolution equations ceases to
satisfy the full Einstein equations as time progresses.

Currently, there are two known sources for constraint violation in an
initial-boundary-value problem. The first one is present already in
the Cauchy problem. It is due to the structure of the field
equations and the specific splitting of these equations into
evolution and constraint equations. The other cause for constraint
divergence is due to inappropriate boundary conditions, i.e., data
given on the boundary of the computational domain which are not
compatible with the constraint equations. These data will give rise to
`constraint violating modes' which propagate into the computational
domain thereby spoiling the solution inside. 

Much work has gone in recent years into the possibilities of curing
the desease of diverging constraints. There have been various
proposals for constraint preserving boundary
conditions~\cite{%
calabreselehner02:_const,%
calabresepullin2003:_well_einst,%
frittelligomez2003:_bound_einst,%
frittelligomez2003:_einst_einst,%
frittelligomez2004:_einst_einst,%
stewart1998:_cauch_ibvp,%
szilagyigomez2000:_cauch,%
szilagyischmidt02:_bound,%
szilagywinicour2003:_well_gener_relat%
} to prevent the constraint violating modes from entering the
computational domain. However, the only formulation of an
initial-boundary-value problem for the Einstein equation which is
known to be well-posed has been given by Friedrich and
Nagy~\cite{friedrichnagy98:_ibvp}. On the other hand, if the
constraints have already started to diverge, there are ways to force
the solution back onto the constraint
surface~\cite{brodbeckfrittelli1999:_equat_asymp_stabl_const_propag,%
lindblomscheel2004:_contr_growth_constr,%
tiglio:_dynam,
holstlindblom:_optim_proj}.

In the present paper we want to discuss the first cause of constraint
violation which is related to the structure of the field equations.
Our aim is to demonstrate that the stability of the constraint
propagation depends heavily on the choice of coordinates, in
particular on the time foliation. We will do this on the basis of an
explicit example, the Bianchi equation. This equation features
prominently in various formulations of the field equations of general
relativity~\cite{%
frauendiener2004:_confor_infin,%
friedrichnagy98:_ibvp,%
friedrich95:_exist_ads%
}
where it can clearly be seen that it is the essential hyperbolic
equation in general relativity. It is the equation which governs the
propagation of the gravitational degrees of freedom described by the
Weyl curvature tensor. However, we want to stress that the method we
employ is general and can be applied to any system of constrained
evolution equations. In fact, a similar analysis can and should be
carried out for the standard Einstein equations in the ADM or BSSN
formulations.

The plan of the paper is as follows. In section~\ref{sec:method} we
present our geometric point of view and discuss our approach in more
detail. In section~\ref{sec:reduction} and~\ref{sec:propagation} we
indicate how to derive the evolution and constraint equations for the
Weyl tensor and how to find the subsidiary system of propagation
equations for the constraints. In section~\ref{sec:stability} we apply the
Routh-Hurwitz criterion for stability to a suitably simplified set of
equations and evaluate the ensuing conditions. Since we are not able
to give complete mathematical proofs for all the statements made, we
dicuss at the end of that section the numerical evidence for our
claims. We end the paper with a brief discussion of the results and
implications for further studies.

\section{General description of the method}
\label{sec:method}

\noindent
In the situation, we are considering, we have to deal with fields on
space-time which constitute a system with infinite dimensions.  In
order to get a feeling for the geometrical situation, we pretend that
we are only concerned with finite dimensions. The material in this
section serves mainly as a motivation for the calculations performed
in the following sections. It is not essential for the rest of paper.
So let $\cP$ denote a finite dimensional manifold which we call the
\emph{phase space} of the system. We assume that there is a vector
field $V$ defined on $\cP$, whose integral curves describe the
evolution in time of the system from some specified initial condition
$p \in \cP$. Thus, $\cP$ can also be interpreted as the manifold of
initial conditions for the system.

Let $\Phi:\cP \to \cZ$ be the \emph{constraint map}, mapping the
initial conditions onto the constraint quantities which form a
manifold $\cZ$. Consider the equation $z=\Phi(p)$ and assume that
there is $p_0 \in \cP$ and $z_0 \in \cZ$ such that $z_0=\Phi(p_0)$.
Let us assume that $d\Phi(p_0)$ is surjective. Then we can locally
write $p=(q,r)$ so that $\del\Phi/\del q$ is invertible. By the
implicit function theorem we can solve the equation $z=\Phi(q,r)$
locally near $p_0=(q_0,r_0)$ i.e., we find a smooth map $q=\psi(z,r)$
with $q_0=\psi(z_0,r_0)$ such that $z=\Phi(\psi(z,r),r)$ for all
$(z,r)$ close to $(z_0,r_0)$.  This allows us to locally consider the
phase space $\cP$ as being parameterised by $p=(\psi(z,r),r)$ where
$z$ are the constraint quantities and $r$ comprises the `residual'
variables. These are the unconstrained `true degrees of freedom'.
Thus, $\cP$ is locally foliated by the leaves of constant $z$.

Consider now the vector field $V$ on $\cP$. Given an arbitrary point
$p_0\in \cP$ there is a unique integral curve $p_t$ of $V$ through
$p_0$ such that $\dot p_t = V(p_t)$. Related to this curve is the
curve $z_t = \Phi(p_t)$ in $\cZ$. This curve describes the change in
the constraint variables caused by the evolution. Its tangent vector
$\dot z_t = d\Phi(p_t)\cdot \dot p_t = d\Phi(p_t)\cdot V(p_t)$ depends
on the solution curve $p_t$. Using the parameterisation for $p_t$ we
can write 
\begin{equation}
\dot z_t = F(z_t,r_t),\label{eq:conpropsys}
\end{equation}
for some smooth map $F:\cP \to T\cZ$.  Now it may happen that for some
$z_0\in \cZ$ we have $F(z_0,r)=0$ for all $r$. Then $z_0=\Phi(p_t)$
for all times $t$ if $z_0=\Phi(p_0)$, i.e., the evolution remains in
the leaf $\cC=\Phi^{-1}(z_0)$, the \emph{constraint manifold}. In that
case, one says that the constraints are propagated by the evolution.
This is the case for the Einstein system in its various formulations
and for many other constrained evolution systems appearing in physics
(however, see~\cite{frauendiener2003:_velo_zwanz} for a well-known
example of a system where this is not the case).

We are interested in the case when the solution curve starts outside
of, but close to, $\cC$ and we want to obtain some information about the
change of $z$ during the evolution. So let us take a 1-parameter
family of evolutions $p_t(\lambda)$  with the corresponding
1-parameter family of constraint variables $z_t(\lambda)$. Assume that
$z_t(0)=z_0$ so that $p_t(0)=(z_0,r_t)$ lies on the constraint surface. Then we
have
\begin{equation}
\frac{\del z_t}{\del t}(\lambda)  = 
F(z_t(\lambda),r_t(\lambda)).\label{eq:conpropsyslambda}
\end{equation}
Taking the derivative with respect to $\lambda$ and evaluating at
$\lambda=0$ yields 
\begin{equation}
  \label{eq:deltaconpropsys}
  \frac{\del}{\del t} \delta z = F_z(z_0,r_t)\cdot \delta z + F_r(z_0,r_t)\cdot
  \delta r = F_z(z_0,r_t)\cdot \delta z.
\end{equation}
The second equality follows by taking the derivative of $F(z_0,r)=0$
with respect to $r$. This equation tells us how perturbations in the
constraint variables close to the constraint surface evolve once they
are excited. Their propagation properties are determined by the
evolution~$p_t$.

In the context of the Einstein equations the evolution curve $p_t$
corresponds to the space-time which evolves from the initial
conditions $p_0$. Thus, the perturbations of the constraint quantities
propagate on the background space-time provided by the solution under
consideration. 

In the finite dimensional setting this is a rather straightforward
route to determine the propagation properties of constraint
perturbations near $\cC$. However, the Einstein system is infinite
dimensional and it is not so clear how much of this route translates
rigorously to an infinite-dimensional setting. In a recent paper,
Bartnik~\cite{bartnik:_phase_einst} shows that much of the
finite-dimensional picture can be taken over to the Einstein system on
asymptotically flat manifolds in the formulation given by Fischer and
Marsden~\cite{fishermarsden1979:_ivp}. In particular, he shows that
the constraint map $\Phi$ is smooth and surjective and that all its
level sets, in particular the constraint manifold $\cC$, are smooth
Hilbert submanifolds of the phase space of GR defined by the first and
second fundamental forms $(g,\pi)$ of a suitable 3-dimensional
manifold. Thus, the infinite dimensional Einstein system shows some of
the features as the finite-dimensional model described above.

The leap from the finite-dimensional to the infinite-dimensional
system is quite considerable. In order to make contact with what
follows let us remark that the map $F$ will in general contain partial
derivatives of the constraint quantities. It is a general procedure to
reduce PDEs to ODEs by Fourier transform.  However, in order to apply
this in the present situation several assumptions need to be made (see
app.~\ref{sec:simpl} ).  Taking the finite-dimensional case as
motivation, we are therefore led to study the linearisation of the
system which propagates the constraints. For a lack of a better name
we call this system following Friedrich the \emph{subsidiary system}.
Since this system is linear in the constraint quantities we have to
study this system itself. In the course of the investigation we may
assume that the background manifold is a fixed solution of the
Einstein equations.  Of course, this procedure is not limited to this
particular formulation of the Einstein equations but applies to any
formulation for which the constraints are propagated by the evolution
equations.

In fact, following Frittelli~\cite{frittelli1997:_note_prop_constr},
Shinkai and Yoneda~\cite{%
shinkaiyoneda:_re_einst,%
shinkaiyoneda2001:_const_adm,%
shinkaiyoneda2002:_adjus_adm,%
shinkaiyoneda2002:_advan_adm,%
yonedashinkai2003:_diagon_const_propag_matric} have already studied
the stability properties of the subsidiary system for several
variations of the ADM system, most notably the BSSN formulation.  Their
work has been motivated by the desire to understand the superiority of
the BSSN scheme over the standard ADM formulation. They analyse
several modified ADM formulations on flat space or on the
Schwarzschild space-time with a fixed time foliation. Compared with
their approach, our work will be both more restrictive and more
general. We do not restrict ourselves to a given background and admit
arbitrary time foliations. But since our aim is to determine the
propagation properties analytically, we cannot easily switch between
various different formulations because the algebra is rather
complicated.

In this paper, we will start with the analysis of the
system of constraint propagation for a class of formulations of the
field equations given by
Friedrich~\cite{friedrich83:_cauch_probl_confor_vacuum_field,%
friedrich95:_exist_ads,%
friedrich96:_hyper_reduct} following this route. These are first order
formulations consisting of equations for several geometrical
quantities, most notably the extrinsic curvature and the acceleration
vector of the time foliation and various curvature quantities
including the Weyl curvature.

The size and resulting complexity of the system makes it hopeless to
analyse the full system at once.  Therefore, we restrict ourselves to
the most important subsystem which features in all these formulations,
namely the so-called Bianchi equation. This is an equation for the
Weyl curvature tensor which results from the Bianchi identity for the
Riemann tensor after using the Einstein vacuum equations. In the
conformal setting there is an additional conformal rescaling involved
which, however, does not change the character of the equation.

As will be described in more detail in the following sections, the
Bianchi equation can be split in the usual way into constraint and
evolution equations, and it can be verified that the constraints
propagate. This collection of constraints and evolution equations will
be referred to as the Weyl system. The subsidiary system of evolution
equations for the constraint quantities contains not only the
constraint quantities of that system but also constraint quantities
whose vanishing implies the consistency of the acceleration and
extrinsic curvature with the existence of a foliation. These
constraints arise when acceleration and extrinsic curvature are also
evolved numerically.  We decouple the Weyl subsystem from the other
subsystems by assuming these constraints to vanish identically i.e.,
that only the perturbations of the Weyl constraints are excited.  This
amounts to the assumption that the Weyl curvature propagates on a
foliation which is not influenced by the curvature and vice versa.

Having obtained the subsidiary system which is already linear in the
constraint quantities, the next step is to localise the equation by
`freezing' the coefficients. This means that we study the system in an
infinitesimal neighbourhood of an arbitrary but fixed event.  This
results in a system with constant coefficients which can be treated by
Fourier analysis. We derive the mode dependent propagation matrix
$P(k)$ and ask for its stability properties. The main tool in this
analysis is the Routh-Hurwitz criterion which allows us to determine
the number of eigenvalues of $P(k)$ with negative real part by looking
at the coefficients of its characteristic polynomial.

One might question the relevance of the frozen problems to the problem
with variable coefficients. This is not an easy task to sort out. One
possibility is to refer to the literature on the analysis of PDEs such
as~\cite{kreisslorenz2004:_initial_bound_value_probl} where it is
shown that for strongly hyperbolic or second order parabolic systems
well-posedness of all frozen systems is sufficient for well-posedness
of the general problem provided there exists a smooth symmetrizer. For
first-order systems Strang~\cite{strang1966:_neces_cauch} has shown
that it is also necessary. This indicates that the properties of the
frozen systems and in particular the estimates which relate the
solution at time $t$ to the initial data are closely related to those
of the general system.




\section{Hyperbolic reduction}
\label{sec:reduction}

\begin{lrbox}{\idbox}
\verb$Id: reduction.tex,v 1.19 2005/07/21 12:51:08 vogel Rel $
\end{lrbox}

\noindent
The formulation of an initial-value problem for the Einstein
equations, which is the basis for their numerical treatment, requires
the introduction of a time-flow along which the integration of the field
equations proceeds to produce a solution out of initial data. The
covariant fields are decomposed into parts tangential and transversal
to this flow (\emph{(3+1)-decomposition}) which splits the
originally covariant field equations into a set of equations for the
(3+1)-constituents of the fields, hopefully yielding a
symmetric-hyperbolic system of evolution equations which allows for
the formulation of a well-posed initial value problem.

In this section we want to present this procedure and the formalism
used here on the example of the Bianchi equation:
\begin{equation}
  \nabla_a °K^a_bcd° = 0
  \eql{bianchi}
\end{equation}
The tensor $°K^a_bcd°$ is a trace-free tensor with the symmetry
properties of the Weyl tensor describing the \emph{gravitational
  field}. Importance and origin of this equation are discussed
in~\cite{penrose65:_zero}.

\subsection{(3+1)-decomposition}
\label{sec:tpo}

We work with the time-like, normalised vector field $t^a$ generating
the time-flow and use metric signature $(+,-,-,-)$, thus $t^at_a =
1$. With respect to this vector field, every tangent space splits into
a parallel (1-dim. time-like) component and an orthogonal
(3-dim. space-like) component. The respective projectors are
\begin{equation}
t^a_b = t^a t_b\ztext{and}
h^a_b = \delta^a_b - t^a t_b\,,
\eql{proj31}
\end{equation}
which also splits the metric:
\begin{equation}
°g_ab° = t_a t_b + °h_ab°\,,
\end{equation}
where $°h_ab°$ is the \emph{negative definite} spatial metric in the
space transversal to $t^a$.

Accordingly, every tensor splits into parts which are parallel or transversal to
$t^a$. We call a tensor purely spatial if every contraction with $t^a$
or $t_a$ vanishes. As an example, the (3+1)-decomposition of $°K^a_bcd°$ is
\begin{align}
  \label{eq:K}
  °K_abcd° &= 4 °t_[a° °E_b][c° °t_d]° + 2 °t_[a°°B_b]^e°°\eps_ecd° - 2 °\eps_ab^e°°B_e[c°°t_d]°
  + °\eps_abe°°E^ef°°\eps_fcd°
\end{align}
with the purely spatial, trace-free and symmetric tensors $°E_ab°$ and
$°B_ab°$ which are called \emph{electric} and \emph{magnetic component
  of the gravitational field}. 
The covariant derivative is decomposed as well:
\begin{equation}
  \nabla_a = t_a D + D_a
  \eql{cd31}
\end{equation}
with its components
\begin{equation}
  D := t^a \nabla_a
  \ztext{and}
  D_a := h_a^b°\nabla_b°\,.
\end{equation}
To facilitate calculations, it is useful to introduce derivative
operators which are adapted to the time vector-field $t^a$ \cite[p.
65]{frauendiener2004:_confor_infin}. To characterise the course of
$t^a$, we introduce the \emph{extrinsic curvature quantities}
\begin{equation}
  \chi^a := D t^a
  \ztext{and}
  °\chi_a^b° := D_a t^b \ntext{(with trace $\chi := °\chi_a^a°$)}\,.
\end{equation}
Since $t^a$ is assumed to be normalised, these quantities are purely
spatial. Usually $t^a$ is chosen as the unit normal field of a
foliation of space-time. Then $°\chi_ab°$ is the \emph{extrinsic
  curvature} of the foliation and \emph{symmetric}. Taking $t^a$ as
the 4-velocity of an observer travelling along the integral curves of
$t^a$, then $\chi^a = t^b \nabla_b t^a$ is the acceleration measured by the
observer. Therefore $\chi^a$ is called \emph{acceleration vector}. The
adapted derivatives are defined by their action on 1-forms:
\begin{align}
  \partial v_b \am:= D v_b + t_b \chi^a v_a - \chi_b t^a v_a\,,
  \eql{del}
  \\
  \partial_a v_b \am:= D_a v_b + t_b °\chi_a^c° v_c - °\chi_ab° t^c v_c
  \eql{dela}
\end{align}
Their action on vector fields and higher tensors is defined by
the Leibniz rule and the requirement that when applied to functions,
they coincide with $D$ and $D_a$.

The adapted derivatives have the important property, that in contrast
to $D$ and $D_a$ they commute with the projectors defined in
\eqr{proj31}. The new time-derivative $\del$ can further be
interpreted as the generator of Fermi-Walker transport along the
integral curves of $t^a$. The spatial derivative $\del_a$ is the
Levi-Civita connection intrinsic to the leaves of the foliation.

\subsection{The Weyl system}

With the mentioned tools, we are in a position to carry out the
(3+1)-decomposition of the Bianchi equation
\begin{align}
  \nabla_a °K^a_bcd° = 0\,.
  \tag*{\eqr{bianchi}}
\end{align}
We first decompose the covariant derivative according to \eqr{cd31},
then transform to the new derivatives $\del$ and $\del_a$ by use of
\eqr{del} and \eqr{dela} and finally insert the (3+1)-representation of
the gravitational field as given by \eqr{K}. The resulting equation
still has three indices but each of its terms can now easily be
classified to be either purely temporal or purely spatial in any of
its indices. This requires every such component of the equation to
hold on its own, thereby splitting the equation into a set of
equations for the (3+1)-components $°E_ab°$ and $°B_ab°$ of the
gravitational field.

The calculations are too technical and lengthy to be given here
\cite{tv04}, therefore we only give the resulting \emph{Weyl system}
of equations:

\paragraph{The constraint equation for $E$:}
\begin{equation}
  \partial^a°E_ac° = °\eps^abe°°\chi_ab°°B_ec° + °\eps_cbe°°\chi_a^b°°B^ae°
  \eql{con_E}
\end{equation}

\paragraph{The constraint equation for $B$:}
\begin{equation}
  \partial^a°B_ac° = - °\eps^abe°°\chi_ab°°E_ec° - °\eps_cbe°°\chi_a^b°°E^ae°
  \eql{con_B}
\end{equation}

\paragraph{The evolution equation for $E$:} 
\begin{equation}
  \partial°E_bc° + °\eps_ae(b°\partial^a°B_c)^e° = - 2 \chi °E_bc° + 2 °\chi^a_(b°°E_c)a°
  + °\chi_(c^a°°E_b)a°-°h_bc°°\chi^ae°°E_ae°+ 2\chi^a°\eps_ae(b°°B_c)^e°
  \eql{ev_E}
\end{equation}

\paragraph{The evolution equation for $B$:} 
\begin{equation}
  \partial°B_bc° - °\eps_ae(b°\partial^a°E_c)^e° = - 2 \chi°B_bc° + 2
  °\chi^a_(b°°B_c)a° + °\chi_(c^a°°B_b)a° -°h_bc°°\chi^ae°°B_ae° 
  - 2°\chi^a°°\eps_ae(b°°E_c)^e°
  \eql{ev_B}
\end{equation}
The system of equations has the remarkable property that it is
invariant under the \emph{duality transformation} $E \ra B$, $B \ra
-E$ known from electrodynamics. This allows for a very compact and
elegant notation: Formally collecting $E$ and $B$ into a complex
tensor $°F_bc° := °E_bc° - i °B_bc°$, the duality transformation now
becomes simply $°F_bc° \ra i \, °F_bc°$. With this, the set of
equations reduces to one single (complex) constraint equation
\begin{align}
    \del^a °F_ac° &
  = i °\eps^abe° °\chi_ab° °F_ec°
  + i °\eps_cbe°°\chi_a^b°°F^ae°
\end{align}
and one single (complex) evolution equation
\begin{align}
  \del °F_bc° + i °\eps_ae(b° \del^a °F_c)^e° &
  = -2\chi°F_bc° + 2°\chi^a_(b°°F_c)a° + °\chi_(c^a°°F_b)a°
   \ct
   - °h_bc°°\chi^ae°°F_ae° + 2i°\chi^a°°\eps_ae(b°°F_c)^e°\,.
\end{align}
It can be shown \cite{friedrich83:_cauch_probl_confor_vacuum_field}
that the evolution equation is symmetric-hyperbolic and that therefore
its initial value problem is well-posed. The development of the
gravitational field is completely determined by the evolution
equation. Thus it defines the vector field $V$ in the picture 
of sect.~\ref{sec:method}.




\section{Constraint propagation}
\label{sec:propagation}

\begin{lrbox}{\idbox}
\verb$Id: propagation.tex,v 1.12 2005/07/21 12:51:08 vogel Rel $
\end{lrbox}

\noindent
The last statement gives rise to the question of compatibility between
the evolution and constraint equations: Given data on an initial time
slice which fulfill the constraint equation, then the evolution
equation fully determines the time development of these data, but will
the constraint equation hold on later time slices as well? In other
words: Is the evolution-generating vector field $V$ tangential to the
constraint surface $\cC$?

This important question can be answered by looking at the time
development of the constraints.  Therefore, we write the constraint
equations \eqr{con_E} and \eqr{con_B} as
\begin{align}
  0 = \E_c
  \am:= \partial^a°E_ac°
  - °\eps^abe°°\chi_ab°°B_ec°
  - °\eps_cbe°°\chi_a^b°°B^ae°,
  \eql{zwang_E}
  \beq
  0 = \B_c
  \am:= \partial^a°B_ac°
  + °\eps^abe°°\chi_ab°°E_ec°
  + °\eps_cbe°°\chi_a^b°°E^ae°
  \eql{zwang_B}
\end{align}
with the \emph{constraint quantities} $\E_c$ and $\B_c$ whose
evolution equations we need to determine. Due to the invariance of the
system of equations under duality transformation we need to calculate
only one of them. The other one is then obtained by substituting $E
\ra B$, $B \ra -E$ and $\E \ra \B$, $\B \ra -\E$.

According to \eqr{zwang_E}, calculating $\del\E_c$ gives on the right
hand side time derivatives of $B$ for which its evolution equation can
be substituted, but furthermore the time derivative of a spatial
divergence of $E$. Using the evolution equation of $E$ in this place
makes it necessary to commute the spatial and time derivative
producing curvature terms. Specialising to the case that $t^a$ is in
fact orthogonal to a foliation and thus assuming $°\chi_ab°$ to be
symmetric, finally yields the following system of propagation
equations:
\begin{align}
  \eql{ev_zwang_E}
    \del\E_c &
  = - \half °\eps_cab°\del^a \B^b
  + \tfrac32 °\eps_cab° \chi^a \B^b
  - \tfrac32 \chi \E_c + \half °\chi_c^b° \E_b 
  \ct
  - °E^ab° °\chiC_cab°
  + 2 °E_ac° °\chiC^ba_b°
  + 2 °\eps_ab(c° °B_d)^b° °\chiC^da°
  \cbeq
  \eql{ev_zwang_B}
  \del\B_c &
  = \half °\eps_cab°\del^a \E^b
  - \tfrac32 °\eps_cab° \chi^a \E^b
  - \tfrac32 \chi \B_c + \half °\chi_c^b° \B_b 
  \ct
  - °B^ab° °\chiC_cab°
  + 2 °B_ac° °\chiC^ba_b°
  - 2 °\eps_ab(c° °E_d)^b° °\chiC^da°
\end{align}
with the constraint quantities of the foliation:
\begin{align}
  \eql{zwang_chi1}
  °\chiC^ab° \am:= °\del^[a°°\chi^b]° \,,\\
  \eql{zwang_chi2}
  °\chiC^abc° \am:= 2 °\del^[a°°\chi^b]c°
  + h^a_{a'} °h^bB° °h^cC° °t^D° °R^A_BCD°
\end{align}

\noindent
Vanishing of $\chiC^{ab}$ guarantees $\chi_{ab}$ to remain symmetric
during evolution, vanishing of $\chiC^{abc}$ is equivalent to the
second Gauss-Codazzi relation which has to hold if $\chi_{ab}$ is the
extrinsic curvature of a foliation.

The first lines in \eqr{ev_zwang_E} and \eqr{ev_zwang_B} feed back the
constraint quantities $\E$ and $\B$ into themselves with the extrinsic
curvature quantities of the foliation acting as coefficients. The
second lines couple the system to the constraint quantities of the
foliation with $E$ and $B$ acting as coefficients.

Obviously all the terms on the right-hand sides are proportional to
constraint quantities. The differential equations therefore are
homogeneous with respect to constraint quantities, i.e. $\E_c = 0$,
$\B_c = 0$ are solutions of the propagation equations under the
assumption, that the constraint quantities of the foliation also
vanish. Since the system can be shown to be symmetric hyperbolic, we
have uniqueness of solutions, so that the only solution with vanishing
initial conditions is in fact the zero solution. Hence it is shown
that the evolution and constraint equations of the Weyl system are
compatible.




\section{Stability analysis}
\label{sec:stability}

\begin{lrbox}{\idbox}
\verb$Id: stability.tex,v 1.27 2005/07/21 12:51:08 vogel Rel $
\end{lrbox}

\noindent
From the analytical point of view, the above statement is all we need.
From the numerical point of view, this is just the first step.

In doing numerics, the canonical approach is to solve the constraint
equations to produce initial data which is \emph{as accurate as
  possible}.  The numerical integration of the evolution equations
will produce a solution of the evolution equations from these initial
conditions. In general, this procedure cannot distinguish between good
data satisfying the constraint equations exactly and bad data
which are perturbed by numerical error. Numerical noise will be
carried along and can (and will) accumulate.

That means that one has to consider the evolution equations from a
more general point of view allowing arbitrary initial data (off the
constraint surface $\cC$). The way in which these non-solutions of the
constraint equations are propagated, depends on properties which are
not part of the original full system of equations but of the evolution
equations \emph{on their own}. 

The form of this system is not fixed.
By adding multiples of the constraint equations one can write down
many different systems with (presumably) very different properties,
changing the evolution anywhere but on the constraint surface $\cC$.
Here we will consider the form of the evolution equations as fixed,
partly because we want to focus on the particular influence of the
foliation and partly because the spinorial formulation of the Weyl
system seems to suggest that this form is a very natural one.

Then we see that the subsidiary
system~(\ref{eq:ev_zwang_E},\ref{eq:ev_zwang_B}) essentially depends
on the foliation chosen for hyperbolic reduction as this is the
parameter which determines how the properties of the covariant
equation~\eqref{eq:bianchi} are partitioned between evolution and
constraint equations.

Since the numerical procedure aims at producing solutions which are as
close to analytic solutions as possible, it must be required to be
stable against perturbations by numerical noise. Thus, it is a
necessary condition that the solutions of the evolution equations
with vanishing constraint quantities are attractors in the positive time
direction, i.e., the constraint surface in the phase space of initial
conditions has to be attractive.

The following analysis will extend the analysis of compatibility given
in the last section, which can be looked upon as stability analysis of
zeroth order, to a stability analysis of first order which is valid
for small perturbations. In this process different approximations have
to be applied. The first one is that we analyse the constraint
propagation properties only within the Weyl system. The coupling to
other equations outside this system will be neglected by assuming the
external constraint quantities (belonging to equations for the
foliation) to vanish.

To make the calculations more compact, we introduce a complex
constraint quantity $\F_c := \E_c - i \B_c$ to exploit the invariance
under duality transformation. Then the \emph{decoupled} propagation
equations combine into a single one which reads:
\begin{align}
  \eql{ev_scrF}
    \del\F_c &
  = \del\E_c - i \del\B_c
  \ceq
  = - \tfrac12 i °\eps_cab° \del^a\F^b + \tfrac32 i °\eps_cab° \chi^a\F^b
  + \tfrac12 °\chi_c^b°\F_b
  - \tfrac32 \chi \F_c
\end{align}
Obviously, $\F_c = 0$ is a solution, but now of particular interest
is, how solutions $\F_c \ne 0$ will behave. If propagation is stable,
they will converge against $\F_c = 0$ in positive time direction. If
not, then the constraint quantity will diverge.

To investigate this, we use another approximation: In general, the
coefficients of the propagation equation $°\chi_a^b°$ and $\chi^a$
vary from point to point. We now consider the propagation properties
\emph{locally} around a point $p\in\mf$ and assume, that in a certain
neighbourhood of this point the coefficients can be considered
constant. That implies, that the space-time manifold is locally
approximated by its tangent space at point $p$. This will be the
manifold of our further investigation. The constraint quantity $\F_c$
and the extrinsic curvature quantities $°\chi_a^b°$ and $\chi^a$
accordingly become tensor fields on the flat tangent space, for
which now is imposed a differential equation with constant
coefficients which formally corresponds to the original equation. The
detailed discussion of what is involved in this step is given in the
appendix~\ref{sec:simpl}. There we show how to derive the final
equation~\eqref{eq:delF_final} which will be analysed here. Note, that
this equation contains the lapse function $N$ and the shift vector.
However, for the purpose here, it is enough to assume $N=1$ and a
vanishing shift, so that $\del=\del_t$. We will comment on the
influence of non-trivial lapse and shift below.

The procedure of this approximation is known as \emph{freezing of
  coefficients}, a standard method in numerical stability analysis.
Because the \emph{frozen} propagation equation is now defined on flat
space, it is apt to be Fourier transformed in the spatial directions.
Let $V$ denote the local spatial tangent space (tangent to the
space-like slice through $p$) and $V^*$ its dual space. Then the
\emph{frozen} constraint quantity can be represented as
\begin{align}
  \eql{pw_scrF}
  \F_c(t,x^a) := \int\limits_{V^*} \Fh_c(t,k_a) e^{i k_a x^a}
  d^3 k_a
  \falle x^a \in V
\end{align}
with its \emph{frequency components} $\Fh_c(t,k_a)$, for which now the
following propagation equation holds:
\begin{align}
    \del\Fh_c &
  =
    \tfrac12 °\eps_cab° k^a \Fh^b + \tfrac32 i °\eps_cab° \chi^a \Fh^b
  - \tfrac32 \chi \Fh_c + \tfrac12 °\chi_c^b° \Fh_b
  \eql{ev_scrFh}
  \ceq
  =
  °\P_c^b°(k_a,°\chi_a^b°,°\chi^a°)\,\Fh_b(t,k_a)
\end{align}

\noindent
The spatial derivative $\del_a$ has transformed into the frequency
covector $k_a$ which reduces the propagation equation to an ordinary
differential equation with constant coefficients which are denoted by
the \emph{propagation tensor}
\begin{equation}
  °\P_c^b° := °\eps_ca^b° \left(\tfrac12 k^a + \tfrac32 i \chi^a\right)
  + \tfrac12 °\chi_c^b° - \tfrac32 \chi h_c^b\,.
  \eql{def_scrP}
\end{equation}

\noindent
The propagation tensor is the evolution generator of the constraint
quantities in frequency representation and its eigenvalues decide on
the propagation properties of the mode belonging to the respective
eigenvalue. 

For numerically stable constraint propagation, the propagation
equation is required to be \emph{stable} and \emph{attractive} around
the point $\Fh_c = 0$. Here \emph{stable} means that the solutions
around $\Fh_c = 0$ can be controlled: For every maximal deviation from
$\Fh_c = 0$ given for all times, there is a maximum initial deviation.
\emph{Attractive} means that in a certain neighbourhood of $\Fh_c = 0$
all solutions converge against $\Fh_c = 0$ for large times.  If both
conditions are met, then the propagation equation is said to be
\emph{asymptotically stable} which is equivalent to all
eigenvalues lying in the left complex half-plane $\{\Re(z) < 0\}$. 
More details can be found in the literature on linear systems, e.g.
\cite{fk_stabil}. The
impact of the location of the eigenvalues and of diagonalisability on
the propagation properties has been analysed by \name{Yoneda} and
\name{Shinkai} in \cite{yonedashinkai2003:_diagon_const_propag_matric}.

To study the eigenvalues, it is necessary to calculate the
characteristic polynomial
\begin{equation}
  \cp{\P}(z) := \det(°\P_a^b° - z h_a^b)\,,
  \eql{cp}
\end{equation}
which is most elegantly done by using the covariant representation of
3-dimensional determinants as
\begin{align}
    \det°T_a^b° &
  = °T_a^[a°°T_b^b°°T_c^c]°\,.
  \eql{det3}
\end{align}

\noindent
Since the propagation tensor is only 3-dimensional in our case, it
would be possible in principle to directly calculate the eigenvalues
by the well known solution formula for the roots of polynomials of
third order. Unfortunately, the solution formula employs case
discriminations which makes the general dependence of the eigenvalues
on the parameters of the propagation tensor difficult to analyse.
Moreover we are only interested in the \emph{sign} of the \emph{real}
part of the eigenvalues to decide whether propagation is stable or
not. Thus, we only need to know under what conditions the spectrum of
the propagation tensor is contained in the left half of the complex
plane. This is exactly the kind of question that can be decided with
the Routh-Hurwitz criterion (see app.~\ref{sec:rh}) which is
applicable to propagation tensors of arbitrary dimensionality.

\subsection{Application to the propagation tensor}
\label{sec:stabilitaet_P}

The first step now is to calculate the characteristic polynomial in
the representation required by the Routh-Hurwitz criterion and results
in
\begin{align}
  \quad \cp{\P}(i z) &
  =
    b_0 z^3 + b_1 z^2 + b_2 z + b_3
    + i\, (a_0 z^3 + a_1 z^2 + a_2 z + a_3)
  \intertext{with the following real coefficients:}
  \eql{rh_koeffizienten}
    a_0 &= 1,\, a_1 = 0 \cnbeq
    a_2 &= - \left(
        \tfrac12 (°\M_a^a°)^2
      - \tfrac12 °\M_b^c°°\M_c^b°
      - \tfrac14 (k_a k^a - 9 \chi_a\chi^a)\right) \cnbeq
    a_3 &= - \tfrac32 k^a °\M_a^b° \chi_b \cnbeq
    b_0 &= 0, \, b_1 = - °\M_a^a° \cnbeq
    b_2 &= - \tfrac32 k_a\chi^a \cnbeq
    b_3 &= \det °\M_a^b°
    - \tfrac14 °\M_a^b° \left(k^ak_b - 9 \chi^a\chi_b\right)
  \intertext{Here, $°\M_ab°$ denotes the symmetric part of the
    propagation tensor which is a trace-transform of the extrinsic
    curvature:}
    °\M_cb° \am:= \tfrac12 °\chi_cb° - \tfrac32 \chi °h_cb°
   \eql{def_scrM}
\end{align}
From these coefficients we calculate the three Hurwitz determinants
$\hd{2}$, $\hd{4}$ and $\hd{6}$ which are required to be strictly
positive for stable constraint propagation. The three inequalities are
of increasing complexity, since $\hd{i}$ is a $i \times
i$-determinant.

\subsubsection{The first Routh-Hurwitz inequality}

The condition requiring $\hd{2}$ to be strictly positive, is
\begin{align}
    0 &< \hd{2} 
  = 
    \begin{vmatrix}
      a_0 & a_1 \\
      b_0 & b_1 \\
    \end{vmatrix}
  = 
    \begin{vmatrix}
      1 & 0 \\
      0 & - °\M_a^a° \\
    \end{vmatrix}
  = - °\M_a^a°
  \underset{\eqr{def_scrP}}{=}
    4 \chi
  \beq
  \gdw
  \chi &> 0
  \tag{RH.1}
  \eql{rh1}
\end{align}
Remarkably, this condition neither depends on the mode $k_a$, nor on
the acceleration vector $\chi^a$. It demands that the foliation has
positive mean curvature at the point under consideration.

\subsubsection{The second Routh-Hurwitz inequality}

This inequality is already much more complicated, therefore we have to
limit ourselves to the results. The complete calculations can be found
in \cite[sec. 6.3]{tv04}. The determinant involved here is:
\begin{align}
  \eql{hd4}
  \hd{4} &
  = 
    \begin{vmatrix}
      a_0 & a_1 & a_2 & a_3 \\
      b_0 & b_1 & b_2 & b_3 \\
      0 & a_0 & a_1 & a_2 \\
      0 & b_0 & b_1 & b_2 \\
    \end{vmatrix}
  \ceq
  = \tfrac12 °\sN_c^c° \left(
      \det°\sN_a^b° - \tfrac14°\sN_a^b° \left(k^a k_b - 9 \chi^a\chi_b\right)
    \right)
    - \left(\tfrac32 k_a\chi^a\right)^2
  \end{align}
  with another trace-transform of the extrinsic curvature
  \begin{align}
    \eql{def_sN}  
  °\sN_a^b° \alignmiddle{:}{=} °\M_a^b° - °\M_c^c° h_a^b
  \us{\eqr{def_scrP}}{=} \tfrac12 °\chi_a^b° + \tfrac52 \chi h_a^b
  \,, 
  \ntext{which implies}
  \beq
  °\sN_a^a° &= °\M_a^a° - 3 °\M_a^a° = -2 °\M_a^a° = 8 \chi\,.
  \eql{sN_trace}
\end{align}
This determinant now contains all the parameters $k_a$, $\chi^a$ and
$°\chi_a^b°$. Ideally one would like to fulfil this condition for
arbitrary modes $k_a$ simultaneously, resulting in relations
between $\chi^a$ and $°\chi_a^b°$ only. Detailed analysis shows that
this is actually possible and yields the following conditions:

Let the acceleration vector $\chi^a$ be represented in polar fashion
as $\chi^a = s\,v^a$ with an unit vector $v^a$ and $s\ge0$.  Then the
following conditions are necessary and sufficient for the second
Routh-Hurwitz inequality $\hd{4} > 0$ to hold for arbitrary
modes $k_a$:
\begin{gather}
  \text{All three eigenvalues $n_i$ of $°\sN_a^b°$ are strictly positive;}
  \eql{rh2}\tag{RH.2}
\end{gather}
for the length $s$ of the acceleration vector hold both
\begin{align}
  s &< \frac23 \sqrt{-\frac{\det°\sN_a^b°}{°\sN_ab° v^a v^b}}
  \eql{rh3}\tag{RH.3}
  \ntext{and}
  \\
  s &\le \frac13 \sqrt{-\frac{\tf12 °\sN_c^c°}{°\isN_ab° v^a v^b}}
  \eql{rh4}\tag{RH.4}\,.
\end{align}
Here $°\isN_a^b°$ denotes the inverse of $°\sN_a^b°$.  The condition
\rh2 implies the first stability condition \rh1. The
conditions \rh3 and \rh4 are not equivalent. Examples show,
that depending on the parameters, either one of them can be more
strict than the other.

\subsubsection{The third Routh-Hurwitz inequality}
\label{sec:hd6}

\hspace{0pt plus 5mm}The third inequality is obtained from the $6\times 6$-determinant 
\begin{align}
    \hd{6} &
  = 
    \begin{vmatrix}
      a_0 & a_1 & a_2 & a_3 &  0  &  0 \\
      b_0 & b_1 & b_2 & b_3 &  0  &  0 \\
       0  & a_0 & a_1 & a_2 & a_3 &  0 \\
       0  & b_0 & b_1 & b_2 & b_3 &  0 \\
       0  &  0  & a_0 & a_1 & a_2 & a_3\\
       0  &  0  & b_0 & b_1 & b_2 & b_3\\
    \end{vmatrix}
    \ceq
  =
    -   °b_1^3°                °a_3^2°
    -   °b_1^2° °a_2^2°                b_3
    +   °b_1^2°  a_2     b_2    a_3
    + 2  b_1     a_2                  °b_3^2°
    - 3  b_1             b_2    a_3    b_3
  \ct
    -            a_2    °b_2^2°        b_3
    +                   °b_2^3° a_3
    -                                 °b_3^3°\,,
\end{align}
with the coefficients already given in \eqr{rh_koeffizienten}.
To examine $\hd{6} > 0$ with respect to the frequency $k_a$, we sort
the terms in $\hd{6}$ by their order in $k_a$:
\begin{align}
  \hd{6} &
  = K_6 + K_4 + K_2 + K_0
\end{align}
The terms $K_i$ contain $k_a$ to the power of $i$ and are given by
\begin{align}
  K_6 \am:= 
    B_1 (
    (B_1 - b_1 A_1)^2
    - °b_2^2° A_1
    )
    \eql{def_k6}
  \cbeq
  \eql{def_k4}
  K_4 \am:=
    ( °B_1° - °b_1° A_1) \bigl(2 b_1 B_1 (A_0+A_2)
    + (B_0+B_2) (A_1 b_1 - 3 B_1)
    + b_1 b_2 a_3 \bigr)
    \ct
    + °b_2^2° B_1 (A_0 + A_2)
    + °b_2^2° A_1 (B_0 + B_2)
    + 2 b_2 a_3 B_1 b_1
    + °b_2^3° a_3
  \cbeq
  \eql{def_k2}
  K_2 \am:=
      °b_1^2° B_1(°A_0° + °A_2°)^2
    + (°B_0° + °B_2°)^2 ( 3 B_1 - 2 A_1 b_1 )
    \ct
    - (A_0 + A_2)(B_0+B_2) \bigl(b_1 (4 B_1 - 2 °b_1° A_1 ) + °b_2^2°\bigr)
    \ct
    + b_2 a_3 °b_1^2° (A_0+ A_2)
    - 3 b_2 a_3 b_1 (B_0+B_2)
    - °b_1^3° °a_3^2°
  \cbeq
  \eql{def_k0}
  K_0 \am:=
    - (B_0 + B_2)\bigl(
    (B_0 + B_2) - b_1 (A_0 + A_2)
    \bigr)^2
  \,,
  \intertext{with the following abbreviations:}
  A_0 \am:= -°\M_a^[a°°\M_b^b]°
  \,,
  A_1 := -\tf14 k^a k_a
  \,,
  A_2 := -\tf94 \chi^a \chi_a\,;
  \non\beq\non
  B_0 \am:= \det°\M_a^b°
  \,,
  B_1 := \tf14 °\M_a^b°k^ak_b
  \,,
  B_2 := \tf94 °\M_a^b°\chi^a\chi_b
\end{align}
This makes it obvious that we have to discuss a polynomial inequality
of sixth order in~$k_a$ of which we hope, that it can be fulfilled for
arbitrary~$k_a$ simultaneously with the first and second Routh-Hurwitz
inequalities above.

To see, if this is actually possible, we will first discuss the low
frequency limit. In the domain of low frequencies, the terms of low
order in $k_a$ play the dominant role. Looking at the case $k_a \ra
0$, only $\hd{6} \ra K_0$ contributes. Analysing this term yields the
following result:

As before let $°\M_a^b°$ denote the symmetric part of the propagation
tensor and choose the acceleration vector $\chi^a$ to be represented
in polar form as $\chi^a = s\,v^a$ with unit vector $v^a$ and length
$s \ge 0$.

Then necessary and sufficient for the third Routh-Hurwitz inequality
$\hd{6} > 0$ to hold in the limit $k_a \ra 0$, are the following
conditions:
\begin{gather}
  \text{All three eigenvalues $m_i$ of $°\M_a^b°$ are strictly negative}\,;
  \tag{RH.5}
  \eql{rh5}
  \\
  s < \frac23 \sqrt{-\frac{\det°\M_a^b°}{°\M_ab° v^a v^b}}
  \tag{RH.6}
  \eql{rh6}\,.
\end{gather}
Because of $n_i = m_i - (m_1 + m_2 + m_3) = -\sum_{j \ne i} m_j$, it
follows from \rh5, that $n_i > 0$ \rh2 which further implies \rh1.
Moreover, it can be
shown, that \rh6 implies the previous conditions \rh3 and
\rh4.  Therefore the conditions \rh5 and \rh6 are
already \emph{sufficient} for the first and second Routh-Hurwitz inequality
$\hd{2},\,\hd{4} > 0$ to hold, and that \emph{for all} $k_a$.

For non-zero frequencies, the inequality $\hd{6} > 0$, of course, is
much more complicated to analyse. Representing the frequency vector in
polar form as $k^a = k\,u^a$ with another unit vector $u^a$ and $k \ge
0$ results in:
\begin{align}
  \hd{6} = K_6(u^a)\,k^6 + K_4(u^a)\,k^4 + K_2(u^a)\,k^2 + K_0(u^a)
  \eql{hd6_polar}
\end{align}
Surprisingly, numerical tests (see below) strongly support the
conjecture, that $K_6(u^a)$, $K_4(u^a)$ and $K_2(u^a)$ are already
\emph{individually} positive whenever the conditions \rh5 and
\rh6 hold. This means, that zero-frequency stability already
implies stability for arbitrary modes.

To test this conjecture in a systematic way, it is useful to
represent the acceleration vector $\chi^a$ in the following
\emph{rescaled} polar form (with \emph{gain} $t\ge0$ and direction
$v_av^a = -1$):
\begin{align}
  \chi^a(t,v^a) \am:= t\,X^a(v^a)\ntext{with}
  \beq
  X^a(v^a) \am:= \frac23\sqrt{-\frac{\det°\M_a^b°}{°\M_ab° v^a v^b}} \, v^a
\end{align}
Then \rh6 is equivalent with $t < 1$.  Inserting this into the
coefficients $K_6$, $K_4$ and $K_2$ of \eqr{hd6_polar} has the
following benefit: Rescaling $°\M_a^b° \ra \alpha\,°\M_a^b°$ now
causes $K_i$ to rescale with a power of $\alpha$: $K_i \ra
\alpha^{9-i}\,K_i$. That means, that rescaling with positive factors
will never change the sign of the individual terms.

Without loss of generality assume that the eigenvalues $m_i$ of
$°\M_a^b°$ are numbered in increasing order: $m_1 \le m_2 \le m_3 <
0$. Then any $°\M_a^b°$ can be represented as $°\M_a^b° = - m_1 \,
°\Mt_a^b°$, where the eigenvalues of $°\Mt_a^b°$ are given by $-1 =
\tilde m_1 \le \tilde m_2 \le \tilde m_3 < 0$ and the $K_i$ will have
the same signs for both tensors with and without tilde. This shows,
that it is sufficient to prove or test the conjecture for eigenvalues
$(-1, \tilde m_2,\tilde m_3)$ lying inside the bounded triangle defined by
the last inequality.

Although we did not find a proof for the given conjecture, we neither
found any special cases in numerical tests, in which the conjecture
would be falsified. For the numerical tests, we used the following
procedure: First, pick the following quantities:
\begin{itemize}
\item $\tilde m_2$, $\tilde m_3$ with $-1 \le \tilde m_2 \le \tilde m_3 < 0$,
\item the gain $t$ with $t < 1$,
\item the direction $v^a$ with $v_a v^a = -1$,
\item the direction $u^a$ with $u_a u^a = -1$.
\end{itemize}
These amount to seven real and independent degrees of freedom, all
bounded to finite ranges. The conditions of the first two items
assert, that the conditions \rh5 and \rh6 hold.

Second, calculate the coefficients $K_6$, $K_4$ and $K_2$ of
\eqr{hd6_polar}. If they are all positive, then the conjecture is
proven for the chosen special case. If $K_6$ is negative, the
conjecture is falsified for the high-frequency limit. If $K_4$ or
$K_2$ are negative, a more detailed analysis must show, if this
results in frequency bands which are instable.

This procedure has been carried out for a large number of special
cases. The eigenvalues have been chosen as $\tilde m_2 := -\alpha$ and
$\tilde m_3 := -\alpha \beta$ with $\alpha$ and $\beta$ in $[10^{-5},1]$. In
different runs we tried both equally distributed and exponentially
distributed values, which concentrate around the critical points
$\alpha = 0$ and $\beta = 0$. The typical $(\alpha,\beta)$-grid had
$20 \times 20$ points.

For the unit vectors $v^a$ and $u^a$ we chose vectors on the unit
sphere in a way, so that the angular distance between them stays
approximately constant for all latitudes, which is not the case for a
simple uniform $(\theta,\phi)$-grid. So the number of points on each
line of latitude increases towards the equator. Further it was taken
advantage of the fact, that the quantities $K_i$ are invariant under
inversion of the unit vectors, so only the northern hemisphere was
actually used. On typical runs, we used $272$ points on this
hemisphere ($10$ latitudes, $40$ longitudes on the equator).

For the choice of the gain $t$, which describes if the condition
\rh6 is met, we also used an exponential spacing, which
concentrates around the critical value $t=1$. We chose $20$ values in
$[0, 1-10^{-5}]$. This amounts to $591.872.000$ combinations per run.

The result of this test is, that in the given domain, we did not find
any cases contradicting the conjecture. Only when choosing values $t
\ge 1$ or $\alpha, \beta \le 0$ we found negative coefficients $K_i$.
As the coefficients $K_i$ only consist of (admittedly complicated)
polynomials, we tend to believe that the conjecture is true.




\section{Results}
\label{sec:results}

\begin{lrbox}{\idbox}
\verb$Id: results.tex,v 1.19 2005/08/12 09:35:29 vogel Rel $
\end{lrbox}

\noindent
In the last section it has been shown that it is possible to apply the
Routh-Hurwitz criterion to the propagation tensor and to perform a
detailed analysis of the geometrical meaning of the resulting
conditions. In our example of the Bianchi equation, this results in:

For the extrinsic curvature $°\chi_ab°$ define the auxiliary tensor
$°\M_ab° := \tf12 °\chi_ab° - \tf32 °\chi_c^c° °h_ab°$.  Represent the
acceleration vector (see \pr{sec:tpo}) in polar fashion as $\chi^a =
s\,v^a$ with unit vector $v^a$ and positive length $s$. Then the
following conditions are \emph{necessary} for \emph{local} stability:
\begin{description}
\item[RH.5]  All eigenvalues $m_i$ of $°\M_a^b°$ are strictly negative;

\item[RH.6] the length $s$ of $\chi^a$ is, depending on its direction $v^a$,
    bounded by
\[  
s < \frac23  \sqrt{-\frac{\det°\M_a^b°}{°\M_ab°v^a v^b}}\,.
\]
\end{description}
Here and in the following \emph{local} stability means asymptotic
stability of the problem which results from localisation in the sense
of the freezing of coefficients approximation.

According to the discussion at the end of \pr{sec:hd6}, we conjecture,
that these conditions are also \emph{sufficient} for \emph{local}
stability. It should be stressed, that these specific conditions of
course only apply to numerical calculations which explicitly integrate
the Weyl subsystem analysed here.

The conditions in a way resemble the partition of the original set of
equations into constraint and evolution equations: As presented in
\pr{sec:tpo}, the extrinsic curvature $°\chi_ab°$ can be defined as
the purely-spatial derivative of the foliation's normal unit vector
field $t^a$, whereas the acceleration vector $\chi^a$ is its temporal
derivative. As \rh5 only contains the extrinsic curvature, it can be
viewed as a \emph{constraint inequality}, because it has to hold on
every individual leaf of the foliation. On the other hand, \rh6
connects the extrinsic curvature with the acceleration vector, thus
forming an \emph{evolution inequality} for the vector field.

We have derived these conditions under the assumption of a constant
lapse and a vanishing shift. It is straightforward to incorporate the
case of non-trivial lapse and shift as follows. Consider the
propagation tensor $\hat{\cP}_a{}^b$ for the general
equation~\eqref{eq:delF_final}. It is easy to verify that
\[
\hat\cP_a{}^b = N\,°\cP_a^b° + i \bigl(N^lk_l\bigr) °h_a^b°.
\]
Thus, the effect of the lapse on the spectrum of $\hat\cP_a{}^b$ is simply
a scaling with a positive number while the shift vector shifts the
spectrum along the imaginary axis. None of these modifications affects
the number of roots in the left half of the complex plane. Therefore,
the stability conditions for the general propagation tensor
$\hat\cP_a{}^b$ are the same as for $\cP_a{}^b$.

One may wonder, why there is no influence of the lapse on the
propagation properties. After all, it is the lapse function which
determines the time-foliation. However, this is easily explained since
it is not the \emph{value} of the lapse which is relevant but its
spatial derivative and this is related to the acceleration vector by
\[
\chi_a^{} = \frac{\del_a N}{N}.
\]
Thus, the spatial variation of $N$ does have a strong influence on the
stability properties.

Results described by \name{Husa} \cite{Husa:2002zc} for Minkowski
evolutions with the conformal vacuum field equations support the
results of our stability analysis. The conformal vacuum equations
given by \name{Friedrich}
\cite{friedrich83:_cauch_probl_confor_vacuum_field} contain the
Bianchi equation explicitly, and when choosing a static hyperboloidal
gauge which satisfies the gauge conditions in the computational
domain, it is observed there that even though the metric components
show exponential divergence from the analytical solution \cite[figure
7]{Husa:2002zc}, the curvature invariants $I$ and $J$ which consist of
components of the Weyl curvature, show exponential \emph{decay}
towards the analytical solution \cite[figure 6]{Husa:2002zc}.

\subsection{Geometric interpretation}

\begin{figure}[hb]
  \begin{center}
    \fbox{a}
    \vspace{1ex}
    
    \includegraphics[height=.4\textwidth]{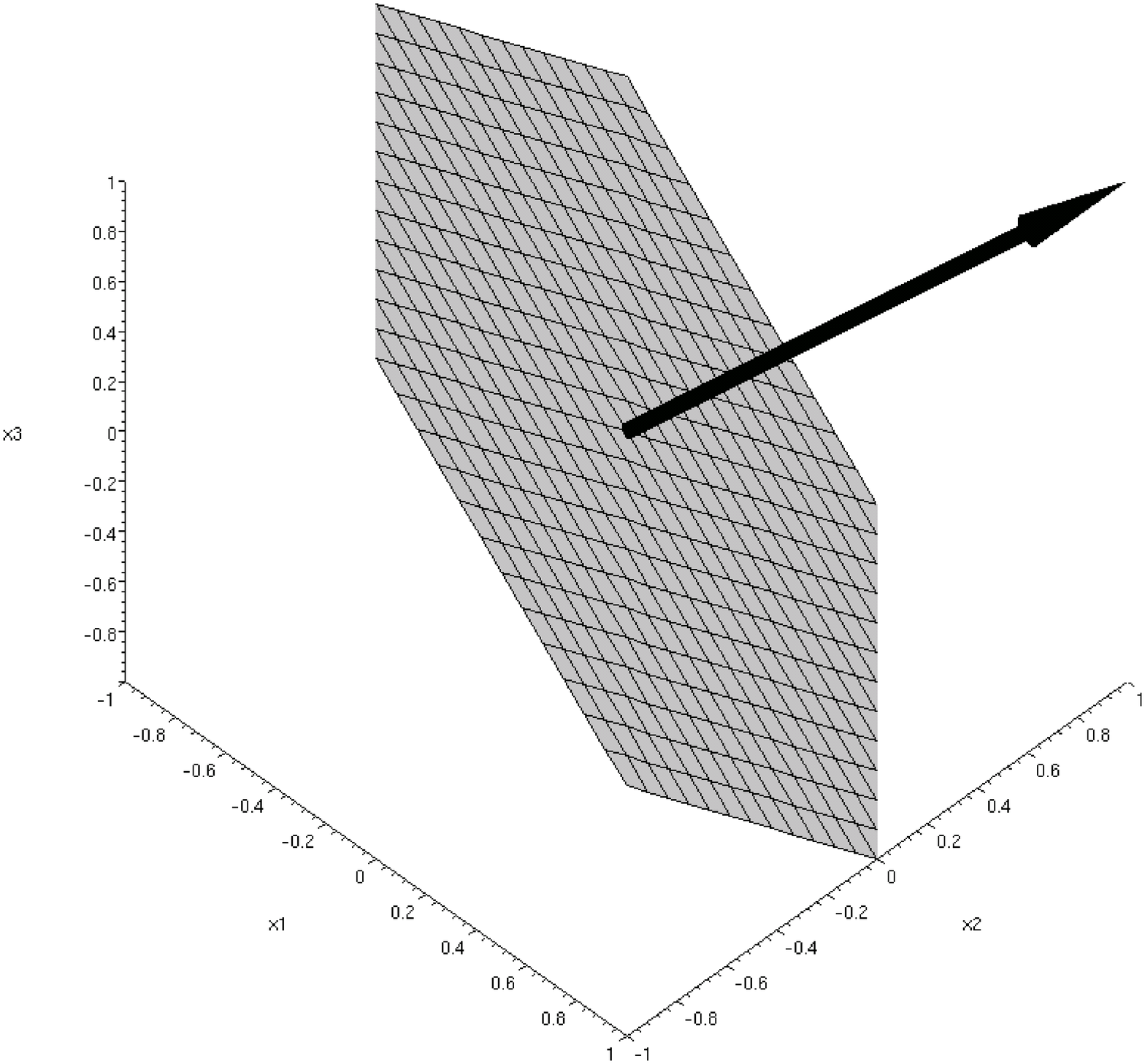}

    \includegraphics[height=.4\textwidth]{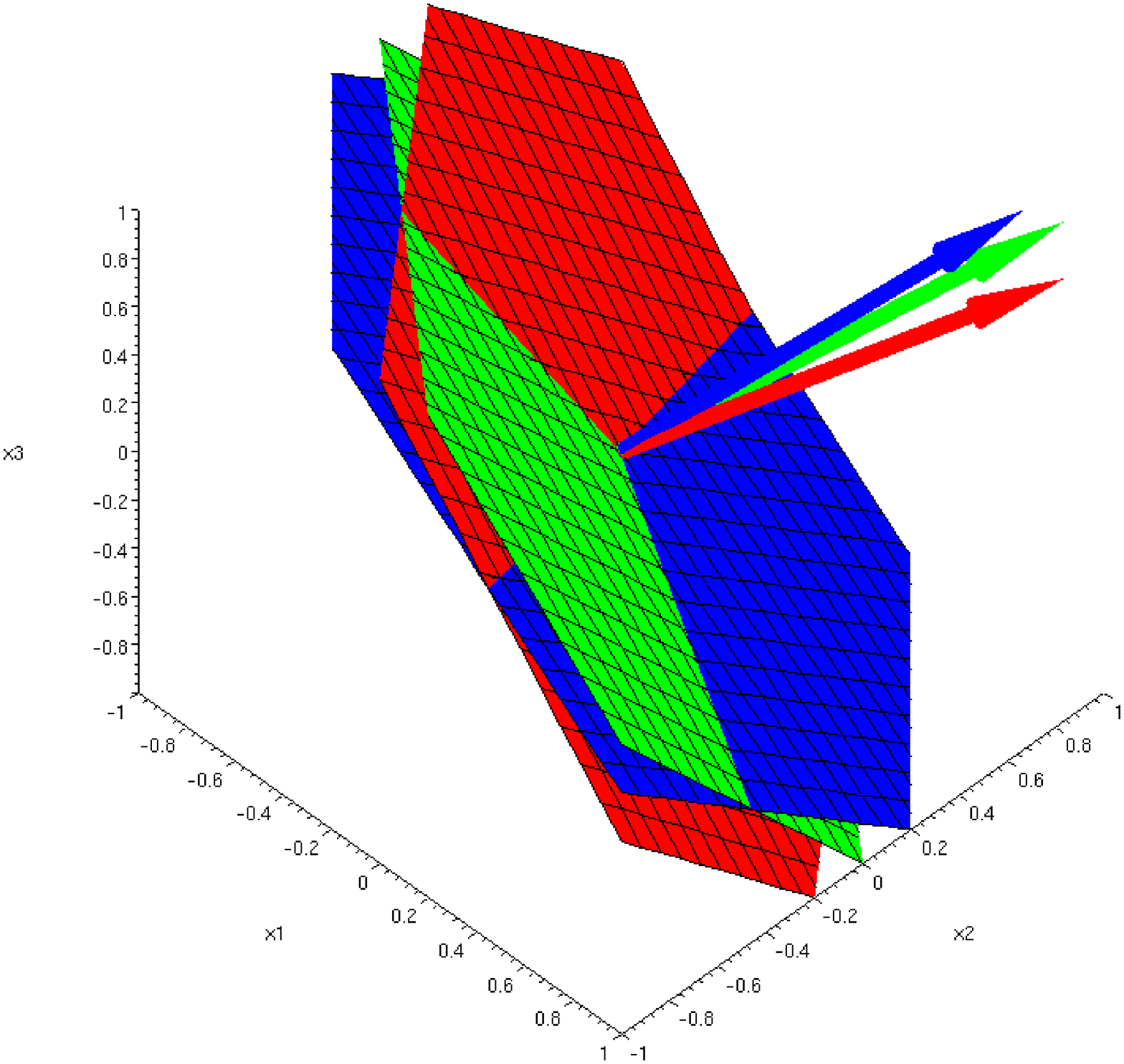}
    \hfill
    \includegraphics[height=.4\textwidth]{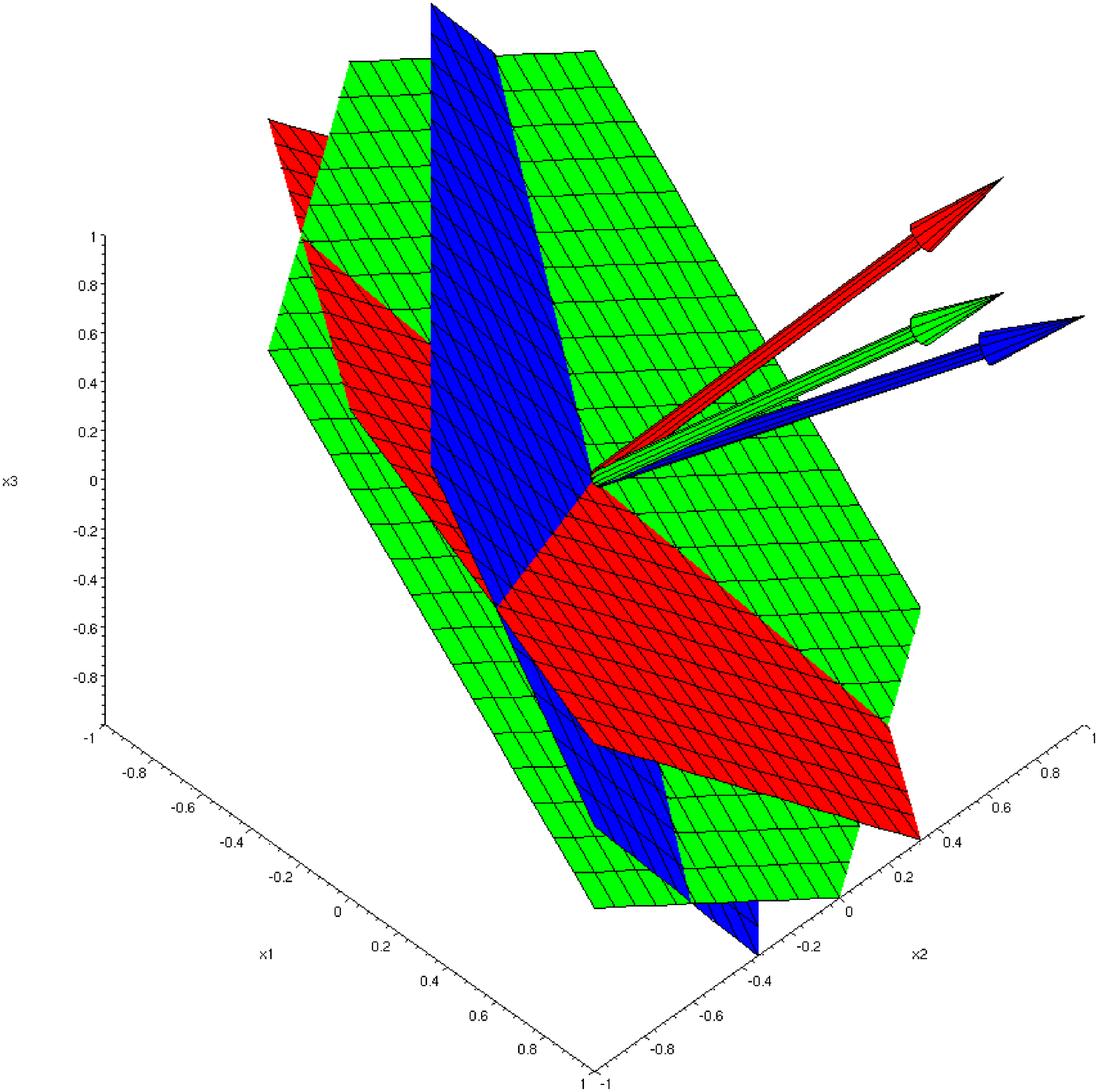}
    \\
    \fbox{b}\hfill \fbox{c}
    \caption{Limitation of the eigenvalues of extrinsic curvature by the conditions 
      a:~$\eqr{rh1}$, b:~$\rh2$ and c:~$\rh5$}
    \label{fig:chi_sN_pos}
  \end{center}
\end{figure}

The conditions \rh1, \rh2 and \rh5 all limit the
allowed eigenvalues of the extrinsic curvature. Their impact is
visualised in \figr{chi_sN_pos}. Each axis in these pictures
corresponds to one eigenvalue of the extrinsic curvature. Then only
such combinations are allowed which lie on the same side of all the
depicted planes as the corresponding arrow. These pictures show that
the conditions are of increasing strictness.

The strongest condition found for the acceleration vector
$\chi^a$ \rh6 can be rewritten as
\begin{align}
    \frac{(\chi^1)^2}{m_2 m_3}
  + \frac{(\chi^2)^2}{m_1 m_3}
  + \frac{(\chi^3)^2}{m_1 m_2} &
  < \frac49
\end{align}
with the components of $\chi^a$ taken with respect to the normalised
eigenvector basis of $°\M_a^b°$. This obviously means, that the vector
$\chi^a$ has to lie strictly inside an ellipsoid whose semi-major axes
are determined by the eigenvalues $m_i$.

\subsection{Trivial foliation of Minkowski space}

As an example consider Minkowski space with standard coordinates
$(t,x,y,z)$. Its flat foliation by $\{ t = \text{const}
\}$-surfaces with vanishing curvature quantities $°\chi_ab° = 0$ and
$\chi^a = 0$ does not fulfil the stability conditions:

Vanishing extrinsic curvature results in $°\M_ab° = 0$ which violates
\rh5. However, this violation is only marginal: The situation
lies exactly on the border of the condition.

Furthermore, the right hand side of \rh6 is undefined. The formal
singularity can be resolved by taking an appropriate limiting
procedure (the numerator is of third order in $\M$, the denominator is
only of first order), but this will still only result in the condition
$s < 0$ which cannot be satisfied. But also here, the condition
\rh6 is only marginally violated.

So it is to be expected, that the eigenvalues of the propagation
tensor lie on the imaginary axis. The propagation equation with
vanishing curvature quantities reads \eqr{ev_scrF}
\begin{align}
  \del °\F_c° = -\frac12\,i\, °\eps_cab° \del^a °\F^b°,
\end{align}
which is completely analogous to the Ampère-Faraday law of vacuum
electrodynamics. The propagation tensor in frequency representation
therefore is $°\P_c^b° = \frac12 °\eps_ca^b° k^a$ with its eigenvalues
$\{0, \pm \frac i2 \sqrt{-k_ak^a}\}$. As expected, their real parts
vanish. Therefore flat Minkowski foliations are not stable in the
strict sense used above, but instead are \emph{marginally stable}.
However, this might be unstable enough to spoil numerical simulations,
since constraint violations will evolve undamped and so can pile up,
even though the growth rate will not be significant.

\pagebreak[4]

\subsection{Hyperboloidal foliation of Minkowski space}

\begin{figure}[htb]
  \centering
  \hspace{0.09\textwidth}%
  \includegraphics[width=0.4\textwidth]{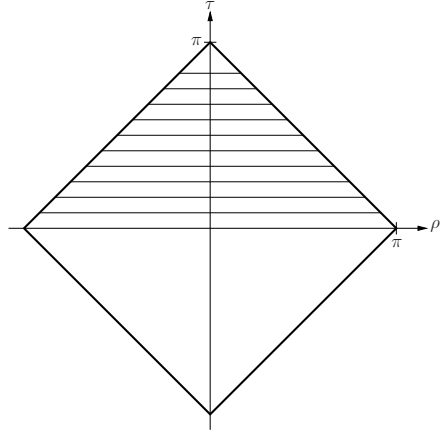}
  \caption{Hyperboloidal foliation of Minkowski space (compactified coordinates)}
  \label{fig:minkowski_compact}
\end{figure}

\begin{figure}[htb]
  \centering
  \includegraphics[width=0.7\textwidth]{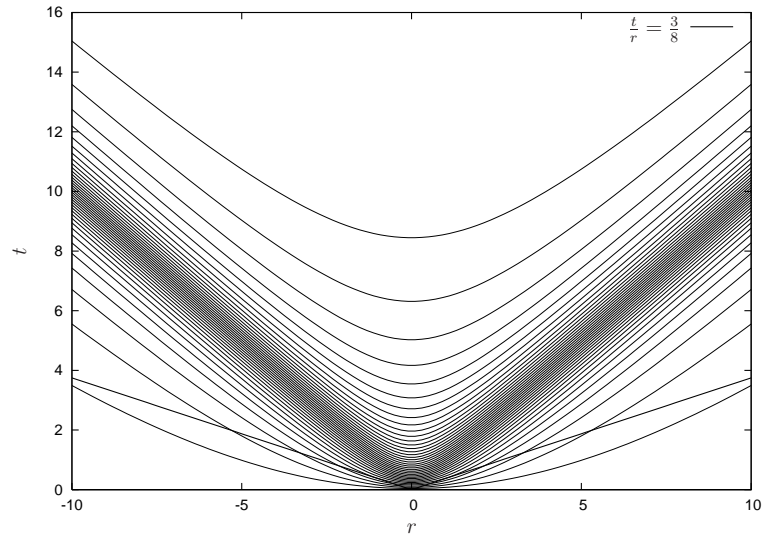}
  \caption{Hyperboloidal foliation of Minkowski space (standard coordinates),
    equally spaced in the compactified time coordinate}
  \label{fig:hyperboloids}
\end{figure}

Now consider the foliation of Minkowski space given by $\{\tau =
\text{const}\}$ surfaces of the parametrisation
\begin{gather}
  t = \frac{\sin \tau}{\cos \tau + \cos \rho}\,,\;
  r = \frac{\sin \rho}{\cos \tau + \cos \rho}
\end{gather}
in spherical coordinates $(t,r,\theta,\phi)$ with compactified time
coordinate $|\tau| < \pi$ and compactified radial coordinate $0 \le
\rho < \pi - |\tau|$ (see \figr{minkowski_compact}). The leaves in
standard coordinates are shown in \figr{hyperboloids}.


This gives for the extrinsic curvature:
\begin{align}
  °\chi_a^b° &= \frac{\sign(\tau)}{\sqrt{\cot^2\tau + 1}} \, h_a^b
  \beq
  \imp
  °\M_a^b° &= - \frac{4\,\sign(\tau)}{\sqrt{\cot^2\tau + 1}} \, h_a^b
\end{align}
That means the condition \rh5 is satisfied only for $\tau > 0$, i.e.
$t > 0$. In this case, the acceleration vector reads
\begin{align}
  \chi^a &= \left[ \frac1{\sqrt{\cot^2\tau + 1}}\,\frac{r}{t} \right] u^a
\end{align}
with the \emph{radial} unit vector $u^a$ tangent to the leaf.
Condition \rh6 is just
\begin{align}
  \frac{t}{r} > \frac38\,,
\end{align}
which means that the condition holds for points which are above
the straight line plotted in \figr{hyperboloids}.




\section{Conclusion}
\label{sec:conclusion}

\begin{lrbox}{\idbox}
\verb$Id: conclusion.tex,v 1.17 2005/08/12 09:35:29 vogel Rel $
\end{lrbox}


\vfill
\noindent
We have shown in this work that the stability of the constraint
propagation is heavily influenced by the choice of the time slicing,
i.e., by the choice of the time coordinate. We derive our results by
an application of the Routh-Hurwitz criterion to the propagation
equation obtained by freezing the coefficients in the system of PDEs
that describes the propagation of the constraints. The system we have
treated is known as the Weyl system because it describes the evolution
of the gravitational degrees of freedom given by the Weyl tensor. We
obtain conditions for the extrinsic curvature and the acceleration
vector of the foliation which have to be satisfied for the local
constraint propagation to be stable.

This work can only be considered as a first step towards understanding
the various contributions to the behaviour of the constraints in
numerical simulations. For instance, we have ignored the effect of
the Ricci rotation coefficients to the propagation tensor by focussing
exclusively on the time foliation.

Furthermore, it is well known that one can add to the \emph{evolution
  equations} terms which are homogeneous in the constraint quantities.
This will generate a subsidiary system whose principal part is
different from the original one and which will have different
stability properties. These possibilities have also been completely
ignored by us. However, general considerations
(see~\cite{brodbeckfrittelli1999:_equat_asymp_stabl_const_propag})
show that such modifications cannot be successful as long as they keep
the time reversal symmetry of the Einstein equations.  Basically, the
argument goes by considering with each initial data set the time
reversed initial data set. If the perturbations of the constraints are
damped in one case then by time reversal, they will grow in the other
case. Thus, the constraint surface cannot be attrative.

We have also treated the Weyl system in isolation, i.e. decoupled from
the rest of the system that describes the time evolution of vacuum
space-times. In this larger system, there is feedback into the Weyl
constraints by the constraints coming from the other subsystems and
vice versa. Again, a more detailed analysis should be made in order
to understand these effects.

Even though we have applied our analysis only to the case of the Weyl
system it is clear that a similar analysis can also be applied to the
standard ADM and the BSSN formulation. Then one could compare the
analytical findings with the study by Shinkai and Yoneda. It would be
interesting to see whether the much improved performance of the BSSN
system compared to the ADM system can be understood from the point of
view of this stability analysis. To end, we quickly recapitulate
the procedure we have followed for our analysis, and which is to be
applied to answer the questions given last:

\begin{enumerate}
\item Starting from the covariant field equations, hyperbolic
  reduction produces a set of \emph{constraint equations} and
  \emph{evolution equations}. The distribution of the properties of
  the covariant equation into the constraint and evolution equations
  is decided by the choice of gauge which appears as coefficients in
  the split equations.
  
\item For the \emph{constraint quantities} whose vanishing indicates
  fulfillment of the constraint equations, \emph{propagation
    equations} can be derived using the evolution equations. These are
  considered as equations for the constraint variables propagating on
  the fixed background provided by some solution of the full system of
  evolution and constraint equations.
  
\item The propagation equations generally couple all of the system's
  constraint quantities. As a first step, feedback of one subsystem
  into itself can be studied by assuming the other constraint
  quantities to vanish which \emph{decouples} the subsystem.
  
\item \emph{Freezing of coefficients} approximates the decoupled
  system \emph{locally} by partial differential equations with
  constant coefficients defined on the local tangential space.
  
\item These \emph{frozen propagation equations} are apt to Fourier
  analysis yielding \emph{propagation equations in frequency
    representation} which are just ordinary differential equations
  with constant coefficients.
  
\item The location of the complex eigenvalues of the matrix of
  coefficients (\emph{propagation tensor}) governs the stability
  behaviour of constraint propagation in this localised picture.
  Necessary for \emph{asymptotic stability} is that the real parts of
  the eigenvalues are negative.
  
\item The \emph{Routh-Hurwitz criterion} is the proper tool to analyse
  the spectrum with respect to stability. It allows to distill
  algebraic conditions for the gauge parameters under which stable
  constraint propagation is possible.
\end{enumerate}

\noindent
These conditions represent necessary and sufficient conditions for
\emph{locally asymptotic} stability of constraint propagation within
the subsystem under analysis. For the stability of the whole system,
they are in general necessary conditions.

\vfill

\section*{Acknowledgment}

\noindent
We are grateful to Prof. H.~P. Hadeler for pointing out the
Routh-Hurwitz criterion to us. This work has been supported by the
SFB-TR7 project on `Gravitational wave astronomy' of the DFG. JF is
grateful to the Erwin-Schrödinger Institut in Vienna for hospitality
while this paper was written. TV is grateful to the SFB 382 `Methods
and algorithms for the simulation of physical processes on super
computers' of the DFG for supporting this research and to the
Max-Planck-Institut für Gravitationsphysik (Albert-Einstein-Institut),
Golm/Potsdam for the opportunity to finish his contribution to this
paper.

\vfill
\vfill


\appendix
\section{Simplifying assumptions}
\label{sec:simpl}

\begin{lrbox}{\idbox}
\verb$Id: simplifications.tex,v 1.6 2005/07/21 12:51:08 vogel Rel $
\end{lrbox}

\noindent
In our treatment of the propagation system it has been necessary to
make some simplifying assumptions in order to bring the equations into
a manageable form. Here, we want to discuss these assumptions in more
detail. Let us start with the covariant form of the propagation
system~\eqref{eq:ev_scrF}
\begin{equation}
    \del\F_c 
  = - \tfrac12 i °\eps_cab° \del^a\F^b + \tfrac32 i °\eps_cab° \chi^a\F^b
  + \tfrac12 °\chi_c^b°\F_b
  - \tfrac32 \chi \F_c.
\end{equation}
This equation is given on the space-time manifold $\mf$ which, we
imagine, has been foliated into space-like leaves $\Sigma_t$ given by
$t=const.$ for some global time coordinate $t$. The time-like
vector field $t^a$ is chosen to be the time-like unit-vector field to
the foliation. In addition, we suppose that three further space-like
unit-vector fields $e^a_i$ ($i=1,2,3$) have been chosen in order to
form a complete tetrad field with $e^a_0=t^a$. The space-like members
of the tetrad are necessarily tangent to the leaves of the foliation.
Let $(t_a=\omega^0_a,\omega^i_a)$ be the dual basis. Then we can write
every tensor field with respect to this basis. Thus we have e.g., 
\begin{equation}
  \label{eq:expansion}
  \F_c = \F_i \omega^i_c.
\end{equation}
Inserting these expansions into the equations and taking components
yields for $l=1,2,3$
\begin{equation}
  \label{eq:delF_expanded}
  \del \F_l - °\Lambda^k_l° \F_k = 
  - i \tfrac12 °\eps_l^mn° \left(\del_m\F_n - °\gamma_m^k_n°\F_k\right)
  + i \tfrac32 °\eps_lmn° \chi^m\F^n 
  + \tfrac12 °\chi_l^m°\F_m
  - \tfrac32 \chi \F_l.
\end{equation}
Here, $\del_k:=e^a_k\del_a$ denotes the directional derivative along
the tetrad vector $e^a_k$ while $\del$ is the directional derivative
along $t^a$. Introducing the familiar (3+1)-split we may write 
\begin{equation}
\frac{\del}{\del t} = N \del + N^k\del_k
\label{eq:split}
\end{equation}
with the lapse function $N$ and the shift vector $N^ke_k^a$.

The functions $°\Lambda^k_l°$ and $°\gamma_m^k_l°$ are Ricci rotation
coefficients with respect to the tetrad defined by
\[
\del e_k^a = °\Lambda^l_k° \,e_l^a \qquad
\del_i e_k^a =  °\gamma_i^l_k°\, e_l^a .
\]
They characterise the behaviour of the spatial tetrad vectors.
Geometrically, the functions $°\Lambda^k_l°$ determine how the spatial
triad is transported from one leaf of the foliation to the next. In
the formulations we are interested in, they are considered as gauge
source functions, i.e., they can be prescribed freely as functions on
$\mf$.  Here, we will assume that they in fact vanish. This amounts to
moving the spatial vectors by Fermi-Walker transport along the
integral curves of $t^a$, the world lines of the Eulerian observers
attached to the foliation (i.e., those observers for which the current
leaf is the manifold of simultaneity).

Now we fix some event $p\in\mf$ which will be the point on which we
localise the equation. Let $t_0=t(p)$ be the value of the time
coordinate at $p$ and let $\Sigma_{t_0}$ be the leaf through $p$. Let
us introduce normal coordinates inside $\Sigma_{t_0}$ centred at $p$
with respect to the metric $h_{ab}$ and choose the spatial triad
\emph{at that point} to agree with the coordinate vectors. Then
\emph{only at the point} $p$ we have $°\gamma_m^k_n°(p)=0$. After
localisation we have~\eqref{eq:delF_expanded} with all the
coefficients being frozen at their value at~$p$. This results in
\begin{equation}
  \label{eq:delF_final}
  \del_t \F_l - N^k\del_k \F_l = \frac{N}{2}\left(
  - i\, °\eps_l^mn° \del_m\F_n 
  + 3 i\,  °\eps_lmn° \chi^m\F^n 
  +  °\chi_l^m°\F_m
  - 3 \chi \F_l\right).  
\end{equation}
Admittedly, this procedure of removing the
functions~$°\gamma_m^k_n°(p)$ is somewhat brutal and not quite
consistent. However, it is the best that we can do if we want to study
the isolated influence of the time foliation on the stability
properties.



\section{The generalised Routh-Hurwitz criterion}
\label{sec:rh}

\begin{lrbox}{\idbox}
\verb$Id: routhhurwitz.tex,v 1.9 2005/07/21 12:51:08 vogel Rel $
\end{lrbox}

\noindent
The appropriate tool to answer questions of stability of ODEs is the
Routh-Hurwitz criterion, also known as Bilharz criterion, which
originates from stability theory. Presentations of the criterion can
be found in \name{Gantmacher} \cite{gant59_2} and \name{Parks} and
\name{Hahn} \cite{ph81}. Our representation follows the lines of
\cite{gant59_2}:

\begin{theorem}[Routh-Hurwitz]
  Let $f$ be a complex polynomial of degree $n$ with
  \[ f(i z) = b_0 z^n + b_1 z^{n-1} + \dotsb + b_n
  + i \left(
    a_0 z^n + a_1 z^{n-1} + \dotsb + a_n
  \right)
  \]
  with $a_i$, $b_i$ real 
  and without loss of generality $a_0 \ne 0$ (otherwise assign $f \ra i f$,
  which does not change its roots). Now define the $2p$-dimensional
  \emph{Hurwitz determinant}
  \begin{equation*}
    \hd{2p} := 
    \begin{vmatrix}
      a_0 & a_1 & \dots & a_{2p-1} \\
      b_0 & b_1 & \dots & b_{2p-1} \\
      0 & a_0 & \dots & a_{2p-2} \\
      0 & b_0 & \dots & b_{2p-2} \\
      \vdots & \vdots & \ddots & \vdots \\
    \end{vmatrix}
  \end{equation*}
  where $p = 1,2,\dotsc,n$ and $a_k = b_k = 0$ for $k>n$.
  Further let both of the following real polynomials
  \begin{align*}
    z \mapsto a_0 z^n &+ a_1 z^{n-1} + \dotsb + a_n \\
    \text{and}\quad z \mapsto b_0 z^n &+ b_1 z^{n-1} + \dotsb + b_n
  \end{align*}
  be coprime, which is equivalent to $\hd{2n} \ne 0$.
  Then the number of roots of the polynomial $f$, which are located in
  the right complex half-plane $\{\Re(z) > 0\}$, is given by
  \begin{equation}
    k = V(1,\hd{2},\hd{4},\dotsc, \hd{2n})\,,
    \eql{rh_k}
  \end{equation}
  with $V(x_1,x_2,\dotsc,x_q)$ denoting the number of changes of sign in
  the sequence $(x_1,x_2,\dotsc,x_q)$. In case some of the determinants
  in \eqr{rh_k} vanish, then for every section of vanishing determinants 
  of length $p$ 
  \[ \hd{2h} \ne 0, \hd{2(h+1)} = \dotsb = \hd{2(h+p)} = 0, \hd{2(h+p+1)} \ne 0
  \quad(h,p \in \N)
  \]
  one has to set:
  \begin{multline}
    V(\hd{2h}, \hd{2(h+1)}, \dotsc, \hd{2(h+p)}, \hd{2(h+p+1)}) \\
    = \left\{
      \begin{array}[c]{ll}
        \frac{p+1}{2} & \text{if p odd} \\
        \frac{p+1-\eps}{2}
        \text{ with } \eps = (-1)^{\frac{p}{2}}\sign\frac{\hd{2(h+p+1)}}{\hd{2h}} &
        \text{if p even}
      \end{array}
    \right.
    \eql{rh_null}
  \end{multline}
\end{theorem}

\noindent
The proof is rather extensive and can be found in the mentioned
literature, the basic idea however will be outlined in the following
section. From the theorem follows, that equivalent to asymptotic
stability is, that all Hurwitz determinants are strictly positive.

\subsection{Mathematical background}
\label{sec:rh_bg}

The basic idea behind the remarkable Routh-Hurwitz criterion is the
\emph{argument principle}: Every polynomial $f$ of degree $n$ has
exactly $n$ complex roots $\alpha_i$ and can therefore be written as a
product of its elementary divisors:
\begin{equation}
  f(z) = \sum_{i=0}^n a_i z^i = a_n \prod_{i=1}^n (z - \alpha_i)
\end{equation}

\noindent
Then the argument of $f$ constitutes additively from the contributions
of the several elementary divisors:
\begin{align}
  \arg f(z) &
  = \arg\left( a_n \prod_{i=1}^n (z - \alpha_i) \right)
  = \sum_{i=1}^n \arg (z-\alpha_i) + \arg a_n    
\end{align}

\begin{figure}[htb]
  \begin{center}
    \includegraphics[height=0.6\textwidth]{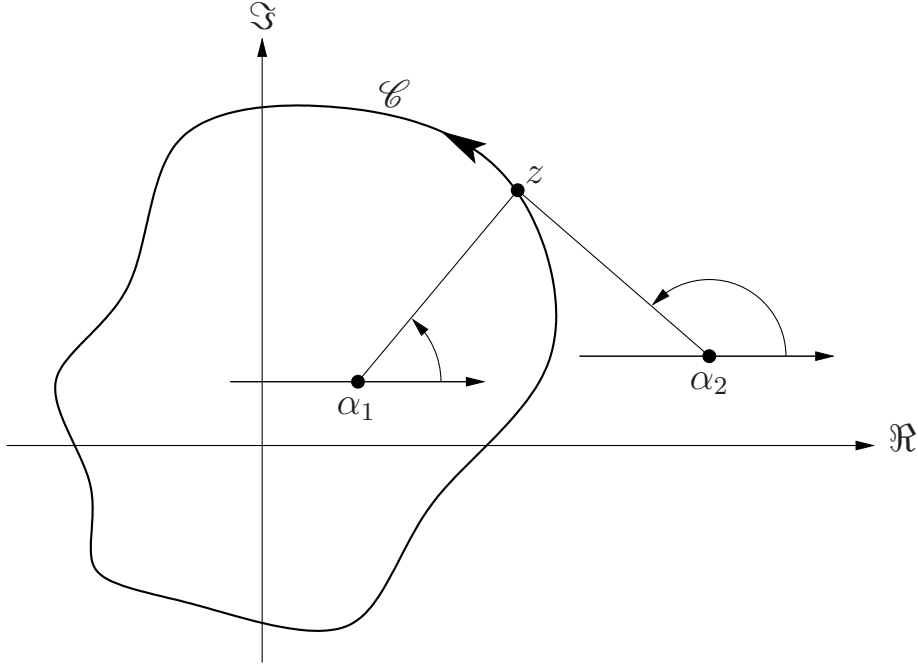}
    \caption{The argument principle}
    \label{fig:argument}
  \end{center}
\end{figure}

\noindent
Now choose a closed, non-self-intersecting path $\pathC$ in the
complex plane and track the change of
$\arg f(z) = \sum_{i=1}^n \arg (z-\alpha_i) + \arg a_n$
while travelling once around $\pathC$ in
positive orientation (s. \figr{argument}).

As the change of argument sums up from the contributions of each root,
one can look at each root individually:

If the considered root lies outside of the domain enclosed by $\pathC$
(e.g. $\alpha_2$ in \figr{argument}), then the argument of $(z -
\alpha_2)$ first grows a little, then decreases and finally increases
so that the net difference exactly vanishes. Roots out of $\pathC$
therefore contribute nothing to the change of argument.

On the contrary, if the root under consideration lies inside $\pathC$
(as $\alpha_1$ of \figr{argument} does), then the argument of
$(z-\alpha_1)$ grows continually, picking up an increase of $2\pi$ for
one revolution.

So the growth of argument counts the number of roots $l$ inside
$\pathC$, counting multiple roots according to their multiplicity:
\begin{equation}
  \label{eq:argument_prinzip}
  \Delta_\pathC \arg f = 2\pi\,l
\end{equation}

\begin{figure}[htb]
  \begin{center}
    \includegraphics[height=0.6\textwidth]{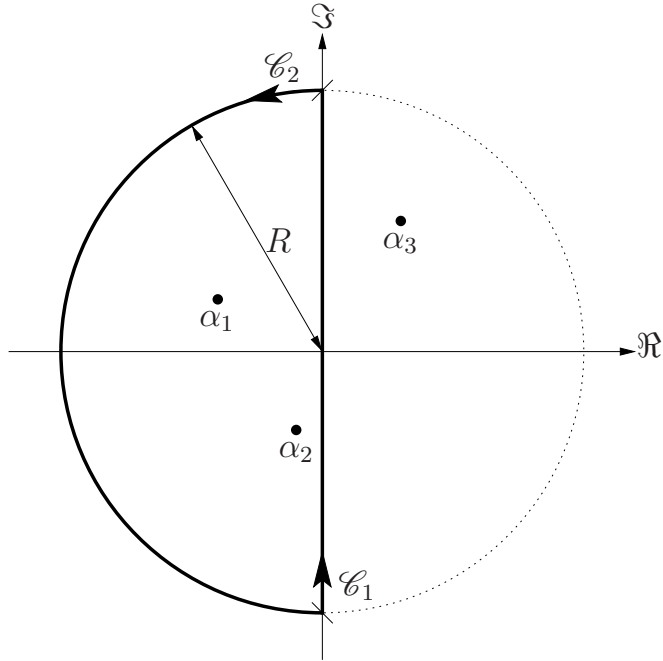}
    \caption{A path enclosing the left complex half-plane for  $R\to\infty$}
    \label{fig:reell_negativ}
  \end{center}
\end{figure}

\noindent
This connection between change of argument and the location of roots
can now be employed to count the roots in a complex half-plane. Let
$l$ and $r$ be the number of roots in the left and right half-plane,
and $n=l+r$, i.e. no roots lie \emph{on} the imaginary axis. Now one
can construct a path $\pathC := \pathC_1 \pathC_2$ out of two segments
(s.  \figr{reell_negativ}), which encloses the left complex half-plane
for $R\to\infty$.

Calculating the change of argument for the individual segments, the
half-circle $\pathC_2$ for $R\to\infty$ always counts all roots,
regardless of their location:
\begin{align}
  \lim_{R\to\infty} \Delta_{\pathC_2} \arg f & =
  \lim_{R\to\infty} \sum_{i=1}^n \Delta_{\pathC_2} \arg
  (z-\alpha_i)
  = \sum_{i=1}^n \lim_{R\to\infty} \left. \arg \bigl( R
    (e^{i\phi} - \tfrac{\alpha_i}{R}) \bigr)
  \right|_{\phi=\tf{\pi}2}^{\tf32\pi}
  \ceq
  = \pi\,n
\end{align}

\noindent
The complete path $\pathC$ however counts only the roots in the left
complex half-plane according to the argument principle:
\begin{align}
  \lim_{R\to\infty} \Delta_{\pathC} \arg f & = 2\pi\,l
\end{align}

\noindent
Thus the contribution of $\pathC_1$ along the imaginary axis is given
as the difference of these two amounts:
\begin{align}
  \lim_{R\to\infty} \Delta_{\pathC_1} \arg f & = \lim_{R\to\infty}
  \Delta_{\pathC} \arg f - \lim_{R\to\infty} \Delta_{\pathC_2} \arg f
  = \pi\,( 2l - n )
  = \pi\,( n-2r )
\end{align}

\noindent
This can be written as
\begin{equation}
  \lim_{R\to\infty} \Delta_{\pathC_1} \arg f = \bigl. \Delta \arg
  f(iz) \bigr|_{z\to-\infty}^{z\to\infty}\,.
\end{equation}

\noindent
Representing the polynomial $f$ by
\begin{equation}
  f(iz) =: f_\Re(z) + i f_\Im(z)
\end{equation}
with real polynomials $f_\Re$ and $f_\Im$, and using
\begin{equation}
  \arg f(iz) = \arctan \frac{f_\Im(z)}{f_\Re(z)}\,,
\end{equation}
now allows to state the \emph{Leonhard-Michailov criterion}:

All roots are located in the left complex half-plane ($r=0$), if and
only if
\begin{equation}
 \arctan \frac{f_\Im(z)}{f_\Re(z)}
\end{equation}
increases exactly by $\pi\,n$ when travelling along the real axis
from $z=-\infty$ to $z=+\infty$.

That means, one has to track the change of angle of the vector
\begin{equation}
  \begin{pmatrix}
    f_\Re(z)\\
    f_\Im(z)
  \end{pmatrix}
\end{equation}
when walking from $z=-\infty$ to $z=+\infty$, and count how often the
vector circles around the origin. This is a rather cumbersome method.
Fortunately, it is possible to find an \emph{algebraic} version of
this criterion. The following shall outline the necessary procedure.
For a proof, we refer to literature, e.g. \cite[section 1.2, p.
10ff]{ph81}.

As a first step, the growth of $\arctan\frac{f_\Im(z)}{f_\Re(z)}$ by
multiples of $\pi$ is directly related to the number of jumps between
$-\infty$ and $+\infty$ of $\frac{f_\Im(z)}{f_\Re(z)}$. For every
increase by $\pi$, the fraction jumps exactly once from $+\infty$ to
$-\infty$.

A magnitude which counts such jumps is the \emph{Cauchy index} $I_a^b
g$ which amounts to the number of jumps of the function $g$ from $-\infty$ to
$+\infty$ minus the number of jumps from $+\infty$ to $-\infty$,
when tracing $g$ from $a$ to $b$.

Considering rational functions $\frac{s_2}{s_1}$, the Cauchy index can
be calculated from the \emph{Sturm sequence} if $s_1$ and $s_2$ are
coprime and the degree of $s_1$ is greater than that of $s_2$. The
Sturm sequence is generated by the \emph{Euclidean algorithm}: For two
elements $s_{i-1}$ and $s_i$ of a Sturm sequence, the next element
$s_{i+1}$ is the negative residual of the polynomial division
$(s_{i-1} : s_i)$, i.e.
\begin{equation}
  s_{i-1} = q_{i-1} s_i - s_{i+1},
\end{equation}
where the quotient polynomials $(q_i)$ are of no further interest.  As
the degree is decreasing from element to element, the Sturm sequence
terminates with a constant polynomial $s_m$ which is non-vanishing if
and only if $s_1$ and $s_2$ are coprime. In this case furthermore the
\emph{Sturm theorem} for the Cauchy index of the rational function holds:
\begin{equation}
  I_a^b \frac{s_2}{s_1} = V(a) - V(b)\,,
\end{equation}
with $V(z)$ denoting the number of changes of sign in the sequence
$\bigl(s_1(z), s_2(z), \cdots, s_m(z)\bigr)$.  Thus one evaluates the
elements of the Sturm sequence at the position $z$ and then counts the
number of changes of sign.

For the Leonhard-Michailov criterion one is interested in the Cauchy
index $I_{-\infty}^{+\infty}\frac{f_\Im}{f_\Re}$, therefore the Sturm
sequence must be evaluated at $z\to-\infty$ and $z\to+\infty$. For
these limiting values only the highest order coefficients in the
elements of the Sturm sequence play a role. They can be expressed in
terms of special determinants, which are composed out of the
coefficients of $f_\Re$ and $f_\Im$.

Finally this leads to the \emph{Routh-Hurwitz criterion} given above, which
therefore represents an algebraic version of the Leonhard-Michailov
criterion.



\providecommand{\bysame}{\leavevmode\hbox to3em{\hrulefill}\thinspace}
\providecommand{\MR}{\relax\ifhmode\unskip\space\fi MR }
\providecommand{\MRhref}[2]{%
  \href{http://www.ams.org/mathscinet-getitem?mr=#1}{#2}
}
\providecommand{\href}[2]{#2}

\end{document}